\pdfoutput=1
\documentclass[longtitle,final,3p,times,twocolumn]{elsarticle}
\usepackage[english]{babel}
\usepackage[utf8]{inputenc}
\usepackage[T1]{fontenc}
\usepackage{lmodern}
\usepackage[italic,endash]{mathastext} %Also change font of mathematics
\usepackage{url}
\usepackage{color}
\usepackage{graphicx}
\usepackage{comment}
\usepackage{amsmath}
\usepackage{amsfonts}
\usepackage{amssymb}
\usepackage{url}
\usepackage{color}
\usepackage{multirow}
\usepackage{threeparttable}
\usepackage[iso]{datetime}
\usepackage{caption} 
\usepackage[alsoload=binary]{siunitx}
\usepackage{relsize}   % Needed by C++ symbol defintion below
\usepackage[version=3]{mhchem}
\usepackage{lineno}
\newcommand*\patchAmsMathEnvironmentForLineno[1]{%
  \expandafter\let\csname old#1\expandafter\endcsname\csname #1\endcsname
  \expandafter\let\csname oldend#1\expandafter\endcsname\csname end#1\endcsname
  \renewenvironment{#1}%
     {\linenomath\csname old#1\endcsname}%
     {\csname oldend#1\endcsname\endlinenomath}}% 
\newcommand*\patchBothAmsMathEnvironmentsForLineno[1]{%
  \patchAmsMathEnvironmentForLineno{#1}%
  \patchAmsMathEnvironmentForLineno{#1*}}%
\AtBeginDocument{%
  \patchBothAmsMathEnvironmentsForLineno{equation}%
  \patchBothAmsMathEnvironmentsForLineno{align}%
  \patchBothAmsMathEnvironmentsForLineno{flalign}%
  \patchBothAmsMathEnvironmentsForLineno{alignat}%
  \patchBothAmsMathEnvironmentsForLineno{gather}%
  \patchBothAmsMathEnvironmentsForLineno{multline}%
}
\usepackage[pdftex, unicode={true}, bookmarks={false}]{hyperref}
\hypersetup{
  colorlinks={true},
  linkcolor={black},
  pdfauthor={
    Johan Nyberg, 
    Department of Physics and Astronomy, 
    Uppsala University},
  pdfcreator={pdflatex},
  pdfkeywords={},
  pdfsubject={},
  pdftitle={},
  urlcolor={blue}
}
\newcommand{\geant}{\ensuremath{\textsc{Geant4}}}
\newcommand{\ROOT}{\ensuremath{\texttt{ROOT}}}
\newcommand{\gfthree}{\ensuremath{\texttt{GF3}}}
\newcommand{\mgt}{\ensuremath{\texttt{mgt}}}
\usepackage{relsize}

\newcommand{\g}{$\gamma$}

\usepackage{soul}

\biboptions{sort&compress}
\journal{Nuclear Instruments and Methods in Physics Research A}

\begin{document}

%%-----------------------------------------------------------------------------

\begin{frontmatter}
  \title{
    Identification and rejection of scattered neutrons in AGATA
    \tnoteref{{t1}}
  }
  \author[ankara]{M.~\c{S}enyi\u{g}it\corref{cor1}}
    \ead{meneksek@science.ankara.edu.tr}
  \author[ankara,kth]{A.~Ata\c{c}}
  \author[sivas]{S.~Akkoyun}
  \author[ankara]{A.~Ka\c{s}ka\c{s}}
  \author[infn-padova,univ-padova]{D.~Bazzacco}
  \author[uppsala]{J.~Nyberg}
  \author[infn-padova,univ-padova]{F.~Recchia}
  \author[infn-milano]{S.~Brambilla}
  \author[univ-milano,infn-milano]{F.~Camera}
  \author[infn-milano]{F.C.L.~Crespi}
  \author[infn-padova,univ-padova]{E.~Farnea\fnref{deceased}}
  \author[infn-milano]{A.~Giaz}
  \author[lnl]{A.~Gottardo}
  \author[surrey]{R.~Kempley}
  \author[csnsm-orsay]{J.~Ljungvall}
  \author[infn-padova,paisley]{D.~Mengoni}
  \author[infn-padova,univ-padova]{C.~Michelagnoli}
  \author[infn-milano]{B.~Million}
  \author[hil]{M.~Palacz}
  \author[univ-milano,infn-milano]{L.~Pellegri}
  \author[univ-milano,infn-milano]{S.~Riboldi}
  \author[lnl]{E.~\c{S}ahin}
  \author[uppsala,riken]{P.A.~S\"{o}derstr\"{o}m}
  \author[lnl]{J.J.~Valiente Dobon}
  \author{and the AGATA collaboration}
  \address[ankara]{Department of Physics, Faculty of Science, Ankara 
    University, TR-06100 Tandoğan, Ankara, Turkey}
  \address[kth]{Deptartment of Physics, Royal Institute of Technology, 
    SE-10691 Stockholm, Sweden}
  \address[sivas]{Faculty of Science, Department of Physics, Cumhuriyet 
    University, 58140 Sivas, Turkey}
  \address[infn-padova]{INFN Sezione di Padova, I-35131 Padova, Italy}
  \address[univ-padova]{Dipartimento di Fisica e Astronomia dell'Universit\`a
    di Padova, I-35131 Padova, Italy }
  \address[uppsala]{Department of Physics and Astronomy, Uppsala University, 
    SE-75120 Uppsala, Sweden}
  \address[infn-milano]{INFN Sezione di Milano, I-20133 Milano, Italy}
  \address[univ-milano]{Universit\'a degli Studi di Milano via Celoria 16,
    20133, Milano, Italy}
  \address[lnl]{Laboratori Nazionali di Legnaro, INFN, I-35020 Legnaro (PD),
    Italy}
  \address[surrey]{Department of Physics, University of Surrey, Guildford,
    GU2 7XH, UK} 
  \address[csnsm-orsay]{Centre de Spectrom\'etrie Nucl\'eaire et de 
    Spectrom\'etrie de Masse - CSNSM, CNRS/IN2P3 and Universit\'e Paris-Sud,
    F-91405 Orsay Campus, France}
  \address[paisley]{Nuclear Physics Research Group, University of
    West of Scotland, High Street, Paisley, PA1 2BE, Scotland, UK}
  \address[hil]{Heavy Ion Laboratory, University of Warsaw, ul. 
    Pasteura 5A, 02-093 Warszawa, Poland}
  \address[riken]{RIKEN Nishina Center, 2-1 Hirosawa, Wako-shi, 
    Saitama 351-0198, Japan}
  \cortext[cor1]{Corresponding author}
  \fntext[deceased]{Deceased} 
  \tnotetext[t1]{In memory of Enrico Farnea}
  \begin{abstract}
    Gamma rays and neutrons, emitted following spontaneous fission of
    \ce{^{252}Cf}, were measured in an AGATA experiment performed at
    INFN Laboratori Nazionali di Legnaro in Italy. The setup consisted
    of four AGATA triple cluster detectors (12 36-fold segmented
    high-purity germanium crystals), placed at a distance of
    \SI{50}{\cm} from the source, and \num{16} HELENA BaF$_{2}$
    detectors.  The aim of the experiment was to study the interaction
    of neutrons in the segmented high-purity germanium detectors of
    AGATA and to investigate the possibility to discriminate neutrons
    and \g\ rays with the \g-ray tracking technique. The BaF$_{2}$
    detectors were used for a time-of-flight measurement, which gave
    an independent discrimination of neutrons and \g\ rays and which
    was used to optimise the \g-ray tracking-based neutron rejection
    methods.  It was found that standard \g-ray tracking, without any
    additional neutron rejection features, eliminates effectively most
    of the interaction points due to recoiling Ge nuclei after elastic
    scattering of neutrons. Standard tracking rejects also a
    significant amount of the events due to inelastic scattering of
    neutrons in the germanium crystals. Further enhancements of the
    neutron rejection was obtained by setting conditions on the
    following quantities, which were evaluated for each event by the
    tracking algorithm: energy of the first and second interaction
    point, difference in the calculated incoming direction of the \g\
    ray, figure-of-merit value.  The experimental results of tracking
    with neutron rejection agree rather well with \geant\ simulations.
  \end{abstract}
\end{frontmatter}

%%-----------------------------------------------------------------------------

%\linenumbers

%%-----------------------------------------------------------------------------

\section{Introduction} \label{s:intro}

Fast neutrons emitted in nuclear reactions can travel long distances
and may deposit part or all of their energies in the sensitive regions
of the detectors, which are used for example for detection of \g\
rays.  An investigation of neutron interactions in \g-ray tracking
spectrometers, like AGATA~\cite{Akkoyun201226} and
GRETA/GRETINA~\cite{Deleplanque1999292,Lee2003,Lee2004}, which are
based on segmented high-purity germanium (HPGe) crystals, is of
special interest for two reasons.  First, an understanding of the
neutron interactions can help to determine, quantify and reduce the
neutron-induced background in the \g-ray spectra. This is the subject
of the present work.  Second, if the neutron interactions can be
detected and identified as such, it may be possible to use the \g-ray
spectrometer also as a neutron detector, with an efficient detection
of the number of emitted neutrons in the nuclear reactions.

In an earlier \geant~\cite{Agostinelli2003250, Farnea2010331}
simulation of the $4\pi$ AGATA \g-ray spectrometer, it was shown that
fast neutrons, with energies between \SI{0.5}{\MeV} and \SI{10}{\MeV},
interact on average \num{7} times with the germanium nuclei in the
AGATA crystals and eventually either escape from the array or become
absorbed through neutron capture~\cite{Ljungvall2005379}.  It was also
estimated that, in the same energy range, the probability of detecting
neutrons by AGATA is about \SI{50}{\percent} assuming a low-energy
threshold of \SI{5}{\keV}. In the same work, the influence of neutron
interactions on the \g-ray tracking algorithms was investigated,
observing that the neutron-induced background reduces the sensitivity
of the spectrometer. This background may become a serious problem in
future experiments with neutron-rich radioactive ion beams, when used
in reactions in which a large number of neutrons are emitted and in
experiments where the cleanness of the \g-ray spectra is of utmost
concern.

Different approaches can be used to discriminate neutrons and \g\ rays
in segmented HPGe detectors, the most obvious one making use of the
time-of-flight (TOF) difference between neutrons and \g\ rays. At the
nominal source to detector distance, which is \SI{23.5}{\cm} for the
AGATA spectrometer, or at shorter distances, this approach is expected
to be of limited use due to the time resolution of the germanium
detectors. Another approach is to make use of the detector
segmentation, either in the \g-ray tracking~\cite{Ljungvall2005379,
  Atac2009554} or pulse-shape analysis procedures. Non-segmented HPGe
detectors did not show any difference in the pulse shapes of neutrons
and \g\ rays~\cite{Ljungvall2005553}. However, segment information may
improve this approach, as suggested by Jenkins et
al.~\cite{Jenkins2009457}, where the inelastic excitation and the
subsequent decay of the $0^{+}$ isomer in \ce{^{72}Ge} was exploited
to tag neutrons by making use of the pulse-shape
analysis. Identification of the neutron interaction points in a
segmented HPGe detector was also investigated by Abt et
al.~\cite{Abt2008}, with the aim of reducing the neutron-induced
background.

In a previous simulation of AGATA, three methods were developed, based
on \g-ray tracking for rejection of the neutron-induced background in
the \g-ray spectra~\cite{Atac2009554}.  These methods made use of the
energy of the first and second interaction point, the incoming
direction of the \g\ ray and the figure-of-merit value of \g-ray
tracking algorithm.  The promising results of the simulations are here
tested with real experimental data obtained with the AGATA
spectrometer. The neutron rejection power of each method is quantified
and different combinations of the methods are investigated in order to
improve the technique.  The experimental setup of this work is
specially designed to get a clear identification of neutrons by
utilising the TOF method, which is achieved by having a large distance
between the source and the AGATA detectors and by using BaF$_2$
detectors as time reference. The TOF results are used to optimise the
neutron rejection methods based on \g-ray tracking.

%%-----------------------------------------------------------------------------

\section{Experiment and data analysis} \label{s:exp}

\begin{figure}[htbp]
  \centering
  \includegraphics[width=0.99\columnwidth]{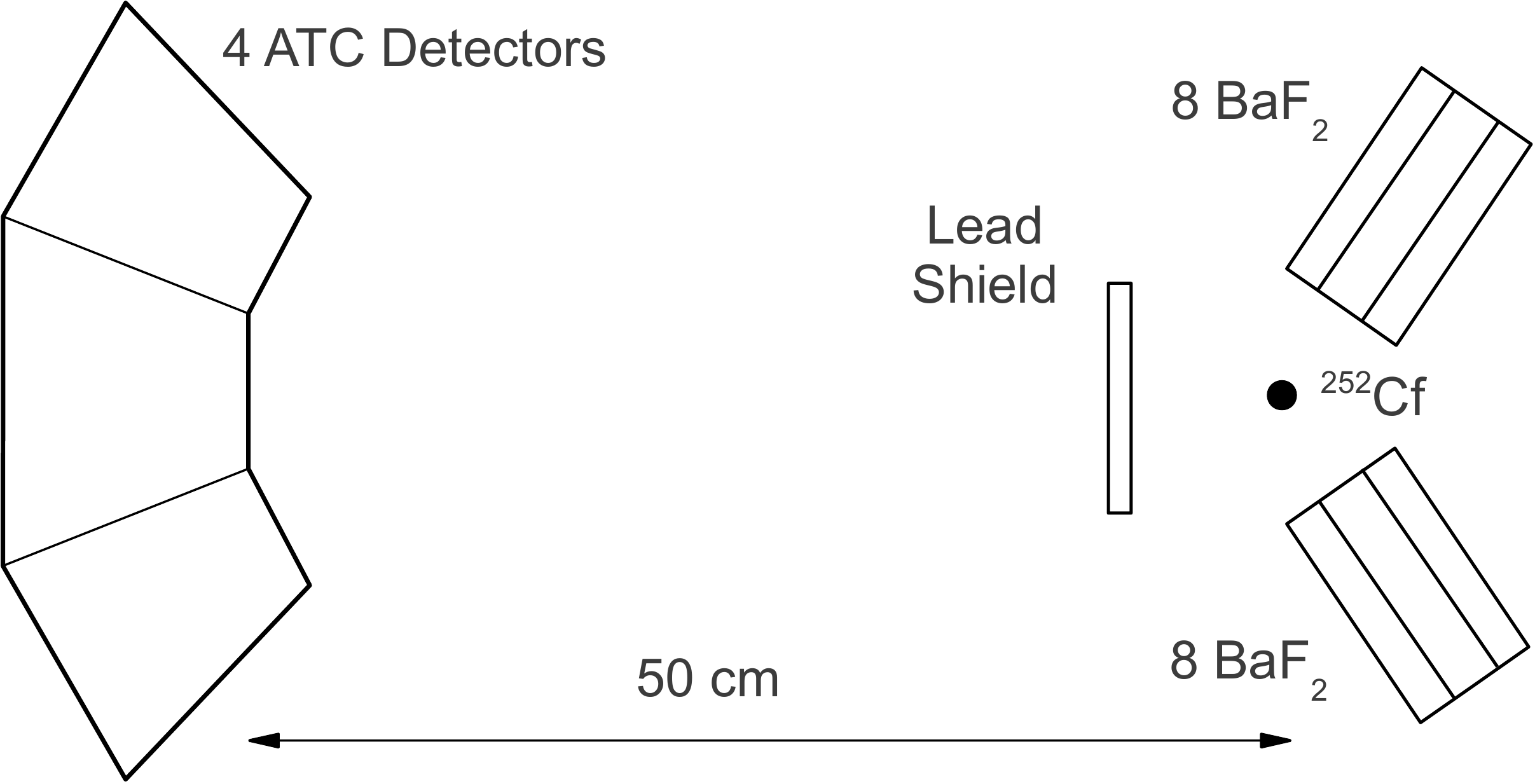}
  \caption{Schematic picture of the experimental setup.}
  \label{f:exp_setup}
\end{figure}

The experiment was carried out during the experimental campaign of
AGATA at INFN Laboratori Nazionali di Legnaro (LNL),
Italy~\cite{Gadea201188, Farnea2012}. A schematic picture of the
experimental setup is shown in Fig.~\ref{f:exp_setup}. A \SI{60}{kBq}
\ce{^{252}Cf} source was placed at a distance of about \SI{50}{\cm}
from the front face of four AGATA triple cluster (ATC)
detectors~\cite{Wiens2010223}, each containing three encapsulated
\num{36}-fold segmented n-type HPGe crystals.  For the measurement of
the TOF of neutrons and \g\ rays, 16 BaF$_{2}$ scintillator detectors
(hexagonal shape, \SI{2}{inch} diameter, \SI{3}{inch} length) from the
HELENA detector array were placed in two groups on each side of the
\ce{^{252}Cf} source at a distance of about \SI{14}{\cm}. In order to
increase the relative yield of neutrons compared to \g\ rays hitting
the ATC detectors, a lead shield with a thickness of \SI{5}{\cm} was
placed between the \ce{^{252}Cf} source and the ATC detectors.  The
shield reduced the \g\ rays and neutrons by about \SI{95}{\percent}
and \SI{50}{\percent}, respectively.

A hardware trigger condition was made in the following way. The
preamplifier core signal of each of the \num{12} HPGe crystals,
obtained from the signal inspection output of the AGATA digitisers,
was fed into an analog timing filter amplifier and further into an
analog constant fraction discriminator (CFD). The CFD thresholds on
the HPGe core signals were set as low as possible, with values ranging
from \SI{10}{\keV} to \SI{20}{\keV}.  A logical OR of the HPGe core
CFD signals was created. The BaF$_2$ anode signals were fed into a
BaF$_2$ signal processor unit~\cite{Boiano2008}, which produced a
logical OR of the CFD signals from each BaF$_2$ detector.  The
hardware trigger signal was made as an AND of the HPGe OR and the
BaF$_2$ OR signals, i.e. at least one HPGe core signal and at least
one BaF$_2$ anode signal was required to produce an event.  The
hardware trigger signal was sent to the AGATA Global Trigger and
Synchronisation (GTS) system via the AGATA VME/VXI interface (AGAVA).

For each HPGe crystal, which contributed to the trigger (core signal
above threshold and in coincidence with HELENA), the pre-processing
electronics produced the following data for readout for the core and
for all \num{36} segments: energy, timestamp, and \num{100} sampling
points (\SI{1000}{\ns}) of the leading edge of the digitised
waveforms.  No low-energy thresholds were applied by the hardware on
the HPGe segment signals: for each crystal, the data from all \num{36}
segment signals were readout whenever a core signal was present.  The
energy and time of each BaF$_2$ detector signal were processed by VME
ADCs and TDCs and readout via the AGAVA interface.  A detailed
description of the AGATA detectors, their electronics and the data
acquisition system can be found elsewhere~\cite{Akkoyun201226}.

The singles HPGe core and BaF$_2$ rates were about \SI{200}{\Hz} and
\SI{500}{\Hz} per detector, respectively, and the trigger rate was
about \SI{120}{\Hz}.  The experiment lasted for about \num{3} days,
during which a total of \SI{5.3e6}{events} were collected.

A \ce{^{60}Co} source calibration, with \num{3} ATC detectors placed
at a distance of \SI{50}{\cm}, was performed before the \ce{^{252}Cf}
experiment. The \ce{^{60}Co} dataset contained about \num{1.0e6}
events and was used to test the influence of the neutron-gamma
discrimination methods on data containing no neutron events.

The three-dimensional positions of the interaction points were
obtained by pulse-shape analysis (PSA) of the recorded signal
waveforms using the adaptive grid-search
algorithm~\cite{Venturelli2005}.  The signal basis used by the PSA
algorithm was obtained from the AGATA Detector Simulation Software ADL
(the version used was from 2010-07-28)~\cite{adl}. It should be noted
that the adaptive grid-search algorithm works very well as long as
there is only one interaction point per crystal. It can also handle
multiple interactions in the same crystal if the transient signals of
the segments do not overlap~\cite{Akkoyun201226}.

In the replay, a digital CFD algorithm was applied on the core signal
waveforms, to calculate the HPGe core time.  The low-energy threshold
applied by this algorithm was about \SI{10}{\keV}.  A TOF parameter
was created as the time difference between the HPGe core time and the
average BaF$_2$ time, obtained as an average of the BaF$_2$ detector
times within a narrow time window.

\subsection{Gamma-ray tracking} \label{ss:tracking}

Gamma-ray tracking was performed by the \mgt\
code~\cite{Bazzacco2004248} that uses the so-called forward tracking
method~\cite{Schmid199969, LopezMartens2004454}, which is based on the
clusterisation of interaction points belonging to the same initial \g\
ray. Among many possible clusters, the tracking code finds and selects
only the one with the best (smallest) figure-of-merit (FM) value. The
FM calculation is based on a comparison of the energy of the scattered
\g\ rays calculated 1) by using the interaction point energies and 2)
by applying the Compton scattering formula to the positions of the
interaction points. The probabilities of photoelectric effect, Compton
scattering and pair production, as well as the probability for the \g\
ray to travel the measured distances between the interaction points in
a cluster, are included in the FM calculation.

The programs \gfthree~\cite{radware} and \ROOT~\cite{root} were used
for the analysis of tracked \g-ray and other histograms.

\subsection{Tracking efficiency and peak-to-total ratio} \label{ss:eff_pt}

A parameter of the \mgt\ code that has an important effect on the
tracking results is the so-called \texttt{lim} parameter, which gives
an upper limit for the FM value. Clusters with larger FM values than
\texttt{lim} are either restructured or rejected.  Depending on the
chosen \texttt{lim} value, either a high tracking efficiency or a
large peak-to-total ($P/T$) ratio is obtained. The selection of the
\texttt{lim} value can be done by using a dataset obtained with a
\ce{^{60}Co} source.  The $P/T$ ratio is defined as the ratio of the
sum of counts in the \SI{1173}{\keV} and \SI{1332}{\keV} peaks to the
sum of counts in the total spectrum, integrated from zero energy or
from a low-energy threshold to just above the \SI{1332}{\keV} peak.
The tracking efficiency is the ratio of the sum of counts of the two
peaks in the tracked \g-ray spectrum to the sum of counts of the peaks
in a spectrum created from the core signals. 

The results obtained from an analysis of the \ce{^{60}Co} dataset, by
using different \texttt{lim} values, are shown in
Table~\ref{t:tab1}. In the present work, \g-ray tracking performed by
using $lim = 0.02$ and none of the tracking methods for neutron
rejection (see
subsection~\ref{sss:tracking_neutron_rejection_methods}), is called
``standard tracking'', since it gives a tracking efficiency and a
$P/T$ value that are closest to what was reported earlier for a setup
with three ATC detectors~\cite{Akkoyun201226}.  Note that for the
largest three \texttt{lim} values shown in the table, basically all
clusters are accepted by \mgt, giving more counts in the \ce{^{60}Co}
peaks as well as increasing the background in the \g-ray spectra.
This explains the large tracking efficiency values, even above
\SI{100}{\percent}, and the small $P/T$ values.

\begin{table*}[htbp]
  \centering
  \caption {The dependency of tracking efficiency and peak-to-total
    on the \texttt{lim} parameter of the \mgt\ tracking program.
    The values listed are for the \SI{1173}{\keV} and \SI{1332}{\keV}
    peaks and were obtained in a measurement with three ATC detectors 
    placed at a distance of \SI{50}{\cm} from the \ce{^{60}Co} source.
    The values in the column ``No threshold'' were obtained by setting
    no additional software threshold on the data, while the values in
    the column ``\SI{200}{\keV} threshold'' were obtained by setting
    a software threshold at \SI{200}{\keV}. To enable a comparison of
    results obtained in previous works~\cite{Akkoyun201226}, 
    only events with more than one interaction point were analysed. 
    The $P/T$ value of a \g-ray spectrum that was created from the 
    core signals was \SI{16.66(7)}{\percent}. No background 
    subtraction was performed. Errors given are purely statistical.}
  \begin{tabular}{|r|c|c|c|}
    \hline
    \multirow{2}{*}{\texttt{lim}} &
    \multirow{2}{*}{Tracking efficiency [\si{\percent}]} &
    \multicolumn{2}{c|}{Peak-to-total ($P/T$) [\si{\percent}]} \\ \cline{3-4}
    & & No threshold & \SI{200}{\keV} threshold \\ 
    \hline
    $10$   & $104.7(8)$ & $41.5(2)$ & $42.2(2)$ \\
    $1$    & $103.3(6)$ & $44.7(2)$ & $45.2(2)$ \\
    $0.5$  & $102.1(6)$ & $46.1(2)$ & $46.6(2)$ \\
    $0.1$  &  $96.8(5)$ & $50.8(2)$ & $51.1(2)$ \\ 
    $0.05$ &  $92.9(5)$ & $53.4(2)$ & $53.6(2)$ \\
    $0.03$ &  $89.0(5)$ & $55.5(3)$ & $55.7(3)$ \\
    $0.02$ &  $85.4(5)$ & $57.4(3)$ & $57.6(3)$ \\
    $0.01$ &  $77.1(4)$ & $60.9(3)$ & $61.1(3)$ \\
    \hline
  \end{tabular}
  \label{t:tab1}
\end{table*}

\subsection{Neutrons and gamma rays emitted by \ce{^{252}Cf}}
\label{ss:cf}

The average prompt neutron and \g-ray multiplicity per \ce{^{252}Cf}
fission are about \num{3.8} and \SI{8}, respectively. The neutrons
have an energy distribution peaking between \SI{0.5}{\MeV} and
\SI{1}{\MeV}, extending to about \SI{10}{\MeV}, and has an average
value of \SI{2.14}{\MeV}. For detailed information about the \g-ray
and neutron emission from \ce{^{252}Cf}, see
e.g. Refs.~\cite{Knoll2010, Valentine2001191, Bleuel2010691,
  Ensslin1991}.

\subsection{Cross sections for neutron capture on \ce{^{nat}Ge}}
\label{ss:cross_sections}

In the neutron energy interval of interest, which is from \SI{1}{\MeV}
to \SI{10}{\MeV}, the total cross section for neutrons on
\ce{^{nat}Ge} is in the range from \SI{3.4}{barn} to
\SI{4.6}{barn}~\cite{Shen2002, Iwamoto2004}.  At {\SI{2.14}{\MeV}} the
total, elastic and inelastic cross sections for neutrons on
{\ce{^{nat}Ge}} are {\SI{3.4}{barn}}, {\SI{2.2}{barn}}, and
{\SI{1.0}{barn}}, respectively~\cite{Shen2002,Iwamoto2004}.  The cross
sections for other reaction channels are much smaller
({\SI{0.2}{barn}}).

After elastic or inelastic scattering of \SI{2}{\MeV} neutrons, the
energy deposited in a HPGe detector by the recoiling Ge nuclei has a
maximum value of about \SI{100}{\keV}~\cite{Atac2009554,
  Ljungvall2005553}.  Due to the pulse-height defect~\cite{Knoll2010},
this energy is reduced to about \SI{35}{\keV}~\cite{Ljungvall2005553}.
In the case of inelastic scattering, both the Ge-recoil energy and
parts or all of the \g-ray energy, originating from the de-excitation
of the excited Ge isotopes, are deposited in the detectors.

\subsection{Distribution of number of interaction points for 
  neutrons and gamma rays} \label{ss:npt}

By using the TOF parameter, the distribution of the number of
interaction points ($npt$) can be obtained separately for neutrons and
\g\ rays. The average $npt$ values obtained for the TOF-gated
distribution was \num{2.35} for neutrons, \num{1.84} for \g\ rays and
\num{1.96} for the un-gated case (see subsection~\ref{ss:ngd_tof} for
a description of the TOF gates).  For the $4\pi$ AGATA array the $npt$
values for neutrons are expected to be significantly larger, while
they will remain essentially unchanged for \g\ rays.

There is a large excess of events with $npt = 1$, which mainly
correspond to elastic scattering of neutrons and small angle Compton
scattering of \g\ rays. Standard tracking rejects about
\SI{75}{\percent} of the single interaction points, since they do not
qualify as photo-absorption points and consequently their FM values
are not good enough. The rest of the single interaction points give
rise to single-hit clusters. The average value of the number of
interaction points in a cluster ($nptc$) after tracking with \mgt\ was
\num{2.78} for neutrons, \num{2.11} for \g\ rays and \num{2.23} for
the un-gated distribution.

\subsection{Identification of gamma-ray peaks} \label{ss:spectra}

\begin{figure*}[htbp]
  \includegraphics[angle=90, width=0.99\textwidth]{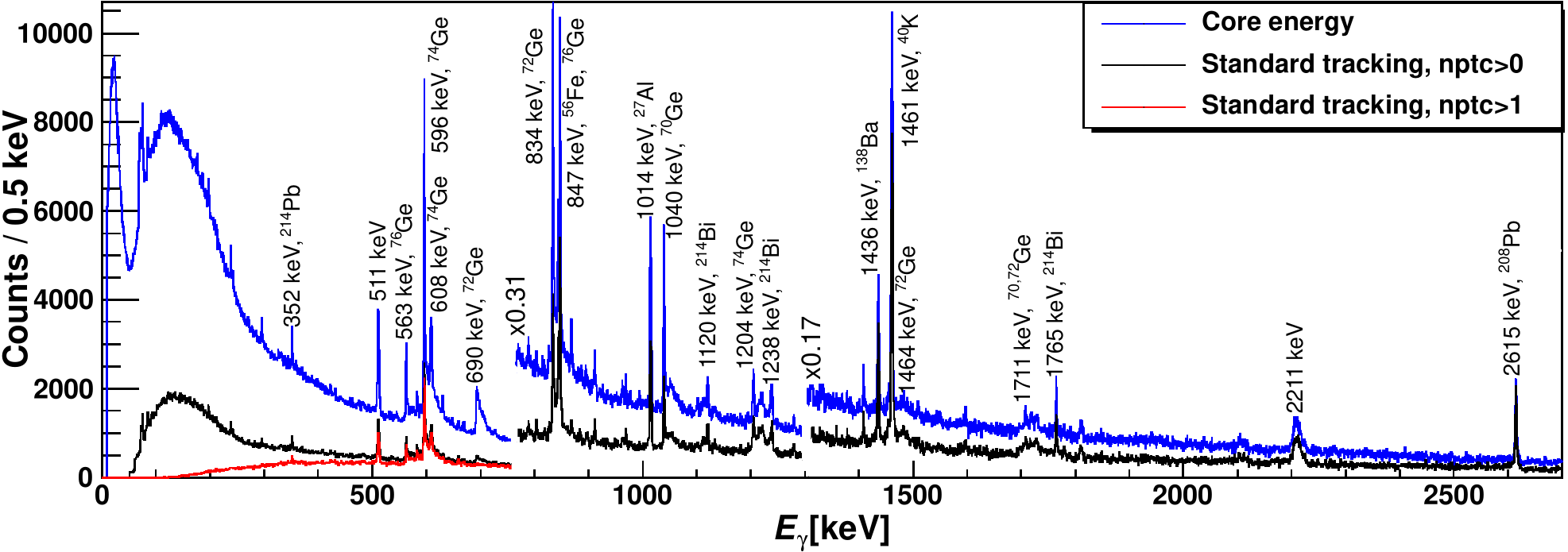}
  \caption{(Colour online) Energy spectra measured with four ATC
    detectors and the \ce{^{252}Cf} source. The black and red spectra
    are obtained after \g-ray tracking including all events ($nptc>0$)
    and including only events with more than one interaction point
    ($nptc>1$), respectively. The blue spectrum is the sum of the core
    energies from the detectors.}
  \label{f:fig2}
\end{figure*}

Due to the \SI{5}{\cm} thick lead block used in the experiment, \g\
rays which originate from the \ce{^{252}Cf} source are highly reduced
in the spectra shown in Fig.~\ref{f:fig2}. The strongest peaks
observed are the ones which originate from the de-excitation of the
different Ge isotopes and from the material surrounding the setup.
Transitions due to the de-excitation of the lowest lying states of
these isotopes are visible both in the core and in the tracked \g-ray
energy spectra.  The \SI{596}{\keV}, \SI{608}{\keV} and
\SI{1204}{\keV} peaks originate from the
$2^{+}_{1}\rightarrow0^{+}_{1}$, $2^{+}_{2}\rightarrow2^{+}_{1}$ and
$2^{+}_{2}\rightarrow0^{+}_{1}$ transitions, respectively, in
\ce{^{74}Ge}.  The \SI{563}{\keV} and \SI{847}{\keV} peaks are due to
the $2^{+}_{1}\rightarrow0^{+}_{1}$ and
$4^{+}_{1}\rightarrow2^{+}_{1}$ transitions, respectively, in
\ce{^{76}Ge}. The \SI{834}{\keV} peak is due to the
$2^{+}_{1}\rightarrow0^{+}_{1}$ transition in \ce{^{72}Ge} and the
\SI{1040}{\keV} peak is due to the $2^{+}_{1}\rightarrow0^{+}_{1}$
transition in \ce{^{70}Ge}. The \SI{690}{\keV} \g\ ray originates from
the de-excitation of the first $0^+$ excited state ($t_{1/2} =
\SI{444}{\ns}$) to the $0^+$ ground state of \ce{^{72}Ge}. Since the
\SI{690}{\keV} peak is due to the detection of conversion electrons,
it is strongly reduced in the tracked \g-ray spectrum (the tracking
algorithm does not include the process of emission of conversion
electrons).

The small bumps observed at \SI{1464}{\keV} and \SI{1711}{\keV} most
likely have contributions from more than one \g-ray transition in
\ce{^{70,72,74}Ge}. It is estimated that the \SI{1464}{\keV} \g\ ray
(\ce{^{72}Ge}) gives the largest contribution to the first bump,
whereas the \SI{1708}{\keV} (\ce{^{70}Ge}) and \SI{1711}{\keV}
(\ce{^{72}Ge}) \g\ rays give the largest contribution to the second
one.

On the high-energy side of the \g-ray peaks, originating from the
\ce{^{70,72,74,76}Ge}(n,n$^{\prime}\gamma$) reactions, about
\SI{40}{\keV} wide bumps, are observed. They appear clearly in the
core-energy spectrum, because the Ge-recoil energies are added to the
energies of the transitions following the
\ce{^{nat}Ge}(n,n$^{\prime}\gamma$) reaction. The bumps are smaller
but still visible in the tracked spectrum. This can be explained
considering that the interaction points due to the Ge recoils are
sometimes included in the \mgt\ clusters and their energies are added
to the peak energies.

The peaks at \SI{847}{\keV}, \SI{1436}{\keV} and \SI{2615}{\keV} are
related to the transitions following the
\ce{^{56}Fe}(n,n$^{\prime}\gamma$),
\ce{^{138}Ba}(n,n$^{\prime}\gamma$) and
\ce{^{208}Pb}(n,n$^{\prime}\gamma$) reactions in the iron material
close to the experimental setup, in the BaF$_{2}$ detectors and in the
lead block, respectively. The peaks at \SI{844}{\keV}, \SI{1014}{\keV}
and \SI{2211}{\keV} are identified as transitions following the
\ce{^{27}Al}(n,n$^{\prime}\gamma$) reaction.  Since the
\SI{2212}{\keV} level in \ce{^{27}Al} has a short mean life
(\SI{26.6}{fs}~\cite{ShamsuzzohaBasunia20111875}), the peak marked at
\SI{2211}{\keV} is Doppler broadened~\cite{Elenkov1982} with a FWHM
value of \SI{15}{\keV}. Due to the natural background radioactivity,
the \ce{^{214}Pb} peak at \SI{352}{\keV}, the \ce{^{214}Bi} peaks at
\SI{1120}{\keV}, \SI{1238}{\keV}, \SI{1765}{\keV} and the \ce{^{40}K}
peak at \SI{1461}{\keV} are observed in this spectrum as a result of
random coincidences.

The counts at low energy, below \SI{45}{\keV}, which are mainly due to
the recoiling Ge nuclei after elastic scattering of neutrons, are
effectively removed from the tracked spectra, as seen in
Fig.~\ref{f:fig2}.  These type of interactions give rise to single-hit
clusters, which are rejected during the \g-ray tracking process (see
subsection~\ref{ss:npt}).  The low-energy bump which appears between
\SI{50}{\keV} and \SI{250}{\keV} in the tracked spectrum originates
from single interaction points with accepted FM values.  The bump
disappears completely in the red spectrum, for which single-hit
clusters with $nptc = 1$ were rejected.

%%-----------------------------------------------------------------------------

\section{Neutron-gamma discrimination} \label{s:ngd}

In this section, the different neutron-gamma discrimination methods
are compared.

\subsection{Time-of-flight} \label{ss:ngd_tof}

A simple and straightforward method for neutron-\g\ discrimination is
to use the difference in TOF of neutrons and \g\ rays. This method
can be successful if the time resolution of the detectors is good
enough and if the variation in flight distance, due to the finite
detector thickness, is small enough for the neutron energy range of
interest.

\begin{figure}[htbp]
  \centering
  \includegraphics[width=1.0\columnwidth]{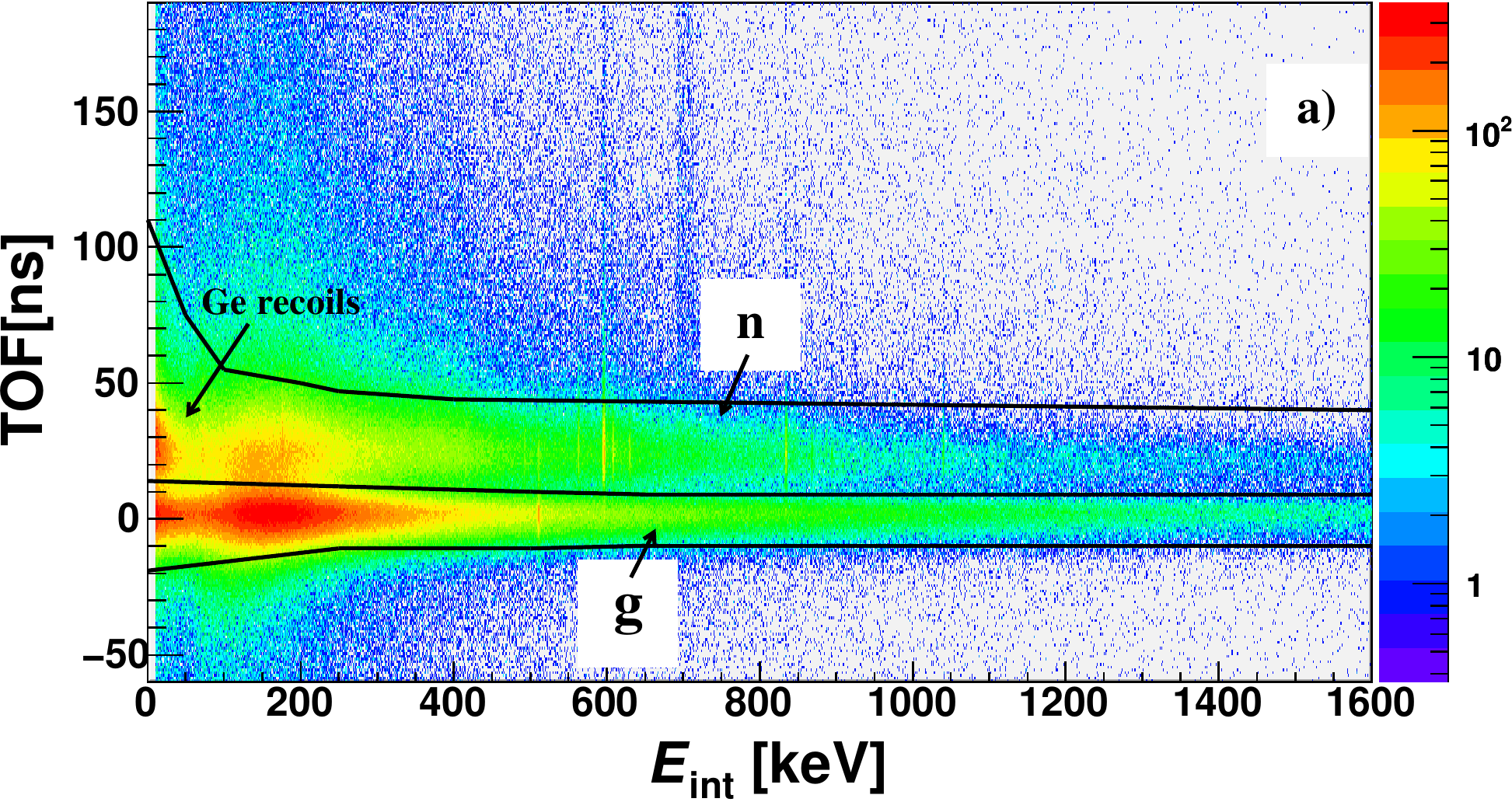}\\
  \includegraphics[width=1.0\columnwidth]{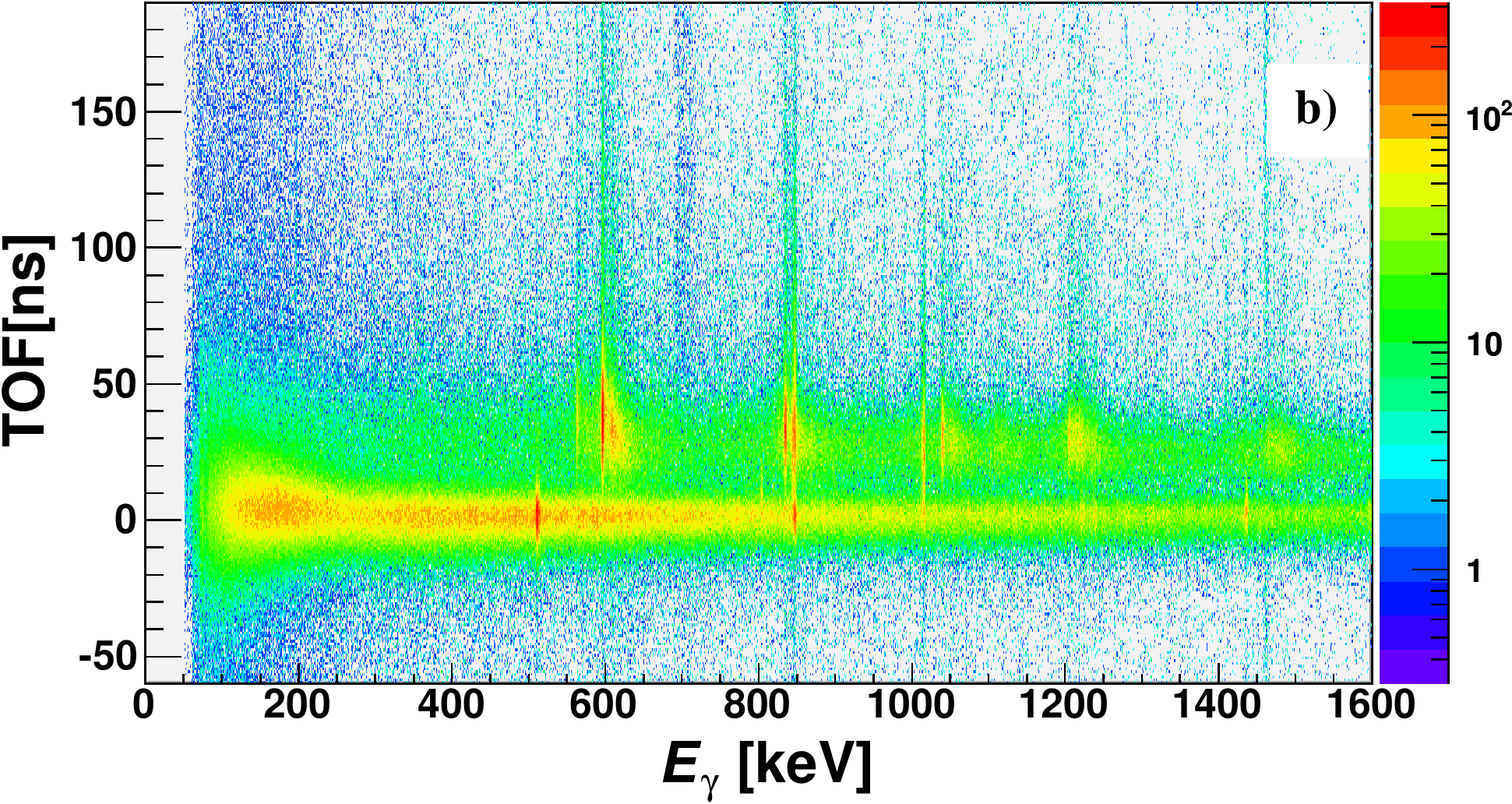}
  \caption{(Colour online) TOF (=time difference between the HPGe core
    and BaF$_{2}$ times) versus a) interaction point energy and b)
    tracked \g-ray energy using standard tracking, measured with the
    \ce{^{252}Cf} source.  The two-dimensional gates shown in a) are
    used for gating on neutrons (n) and \g\ rays (g).}
  \label{f:fig3}
\end{figure}

The histograms in Fig.~\ref{f:fig3} show a) TOF as a function of the
energy of the interaction points deposited in the Ge crystals and b)
TOF as a function of the tracked \g-ray energy using standard
tracking. In Fig.~\ref{f:fig3}a, \g\ rays due to inelastic scattering
of neutrons on different Ge isotopes can be seen at \SI{596}{\keV}
(\ce{^{74}Ge}) and at \SI{834}{\keV} (\ce{^{72}Ge}) for TOF values
larger than about \SI{10}{\ns}. In the region of low energies
($E_{\textnormal{int}} \lesssim \SI{45}{\keV}$) and large TOF values
($\gtrsim \SI{10}{\ns}$), interaction points due to recoiling Ge
nuclei after elastic and inelastic scattering of neutrons can be
seen. The \SI{690}{\keV} \g\ ray (\ce{^{72}Ge}), which is due to the
de-excitation of the $0^+$ first excited state with a half life of
\SI{444}{\ns}, is visible as a broad structure at large TOF values.

In Fig.~\ref{f:fig3}b, the \SI{596}{\keV}, \SI{834}{\keV},
\SI{1040}{\keV}, \SI{1204}{\keV}, \SI{1464}{\keV} and \SI{1708}{\keV}
\g\ rays, which follow the \ce{^{nat}Ge}(n,n$^{\prime}\gamma$)
reactions, are observed at TOF values larger than about \SI{10}{\ns},
together with their bumps on the high-energy side of the peaks. The
\SI{1014}{\keV} \g\ ray, following the
\ce{^{27}Al}(n,n$^{\prime}\gamma$) reaction, is observed both at small
and large TOF values.  Aluminium can be found in a number of places in
the experimental area both at short and long distances from the
\ce{^{252}Cf} source, e.g. in the detector capsules, which leads to
large distances and long neutron flight times, and in the material
that is located close to the \ce{^{252}Cf} source, leading to short
flight times.  The \SI{1436}{\keV} \g\ ray, which is emitted following
the \ce{^{138}Ba}(n,n$^{\prime}\gamma$) reaction, is visible at TOF
values less than about \SI{20}{\ns}. This is expected due to the
relatively short neutron flight distance for reaching the BaF$_{2}$
detectors.  Finally, the \SI{511}{\keV} peak, mainly originating from
pair-production and positron annihilation in the HPGe crystals, is
clearly visible in Fig.~\ref{f:fig3}b.

\begin{figure}[htbp]
  \centering
  \includegraphics[width=0.99\columnwidth]{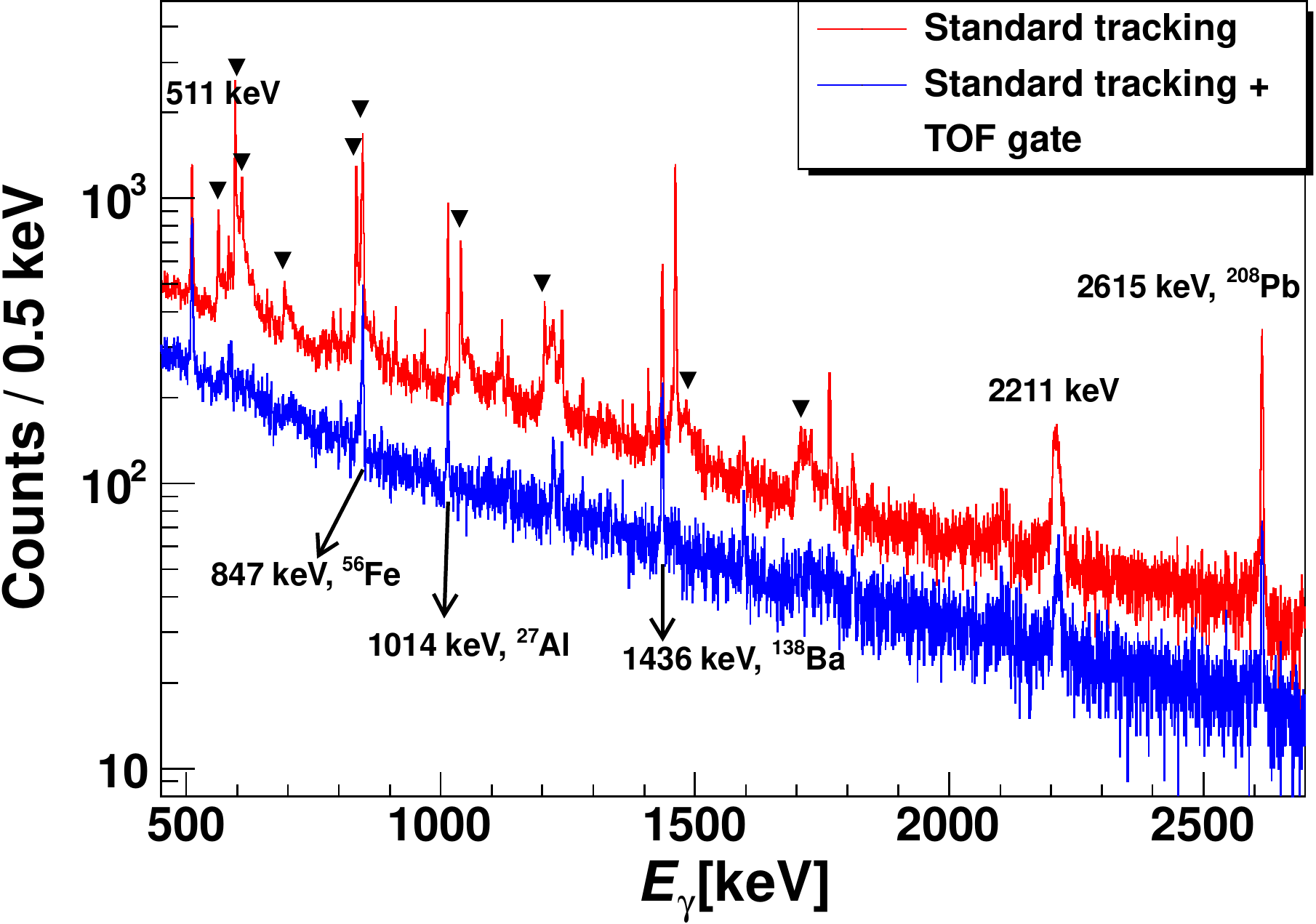}\\
  \caption{(Colour online) Gamma-ray energy spectra obtained with the
    \ce{^{252}Cf} source and by using standard \g-ray tracking without
    any TOF gate (red) and with a TOF gate on \g\ rays (blue; gate g
    in Fig.~\ref{f:fig3}a).  The \SI{511}{\keV}, \SI{847}{\keV},
    \SI{1014}{\keV}, \SI{1436}{\keV}, \SI{2212}{\keV} and
    \SI{2615}{\keV} \g\ rays are emitted following positron
    annihilation and the reactions \ce{^{56}Fe}(n,n$^{\prime}\gamma$),
    \ce{^{27}Al}(n,n$^{\prime}\gamma$),
    \ce{^{138}Ba}(n,n$^{\prime}\gamma$),
    \ce{^{27}Al}(n,n$^{\prime}\gamma$) and
    \ce{^{208}Pb}(n,n$^{\prime}\gamma$), respectively.  The black
    triangles indicate \g\ rays due to inelastic scattering of
    neutrons on the stable Ge isotopes.  The \g-gated spectrum (blue)
    contains \SI{68}{\percent} of the counts in the un-gated spectrum
    (red).}
  \label{f:fig4}
\end{figure}

Two-dimensional TOF gates, indicated by n (neutrons) and g (\g\ rays)
in Fig.~\ref{f:fig3}a, were defined in order to make a TOF gated
neutron-\g\ discrimination that was as clean as possible.  For an
\mgt\ cluster to be qualified as a \g\ ray, all of its interaction
points were required to be in the g gate.  Fig.~\ref{f:fig4} shows
tracked \g-ray spectra with no TOF gate and with a TOF gate on \g\
rays (gate g).  It is clear from this figure that standard tracking
alone cannot eliminate the neutron-induced peaks and and their
bumps. They are, however, effectively eliminated by using the TOF
gate. As expected, the peaks at \SI{511}{\keV}, \SI{847}{\keV},
\SI{1014}{\keV}, \SI{1436}{\keV}, \SI{2211}{\keV} and \SI{2615}{\keV},
which originate from positron annihilations and from the reactions
\ce{^{56}Fe}(n,n$^{\prime}\gamma$),
\ce{^{27}Al}(n,n$^{\prime}\gamma$),
\ce{^{138}Ba}(n,n$^{\prime}\gamma$),
\ce{^{27}Al}(n,n$^{\prime}\gamma$) and
\ce{^{208}Pb}(n,n$^{\prime}\gamma$), respectively, are still visible
after the TOF gating.

\begin{figure}[htbp]
  \centering
  \includegraphics[width=0.99\columnwidth]{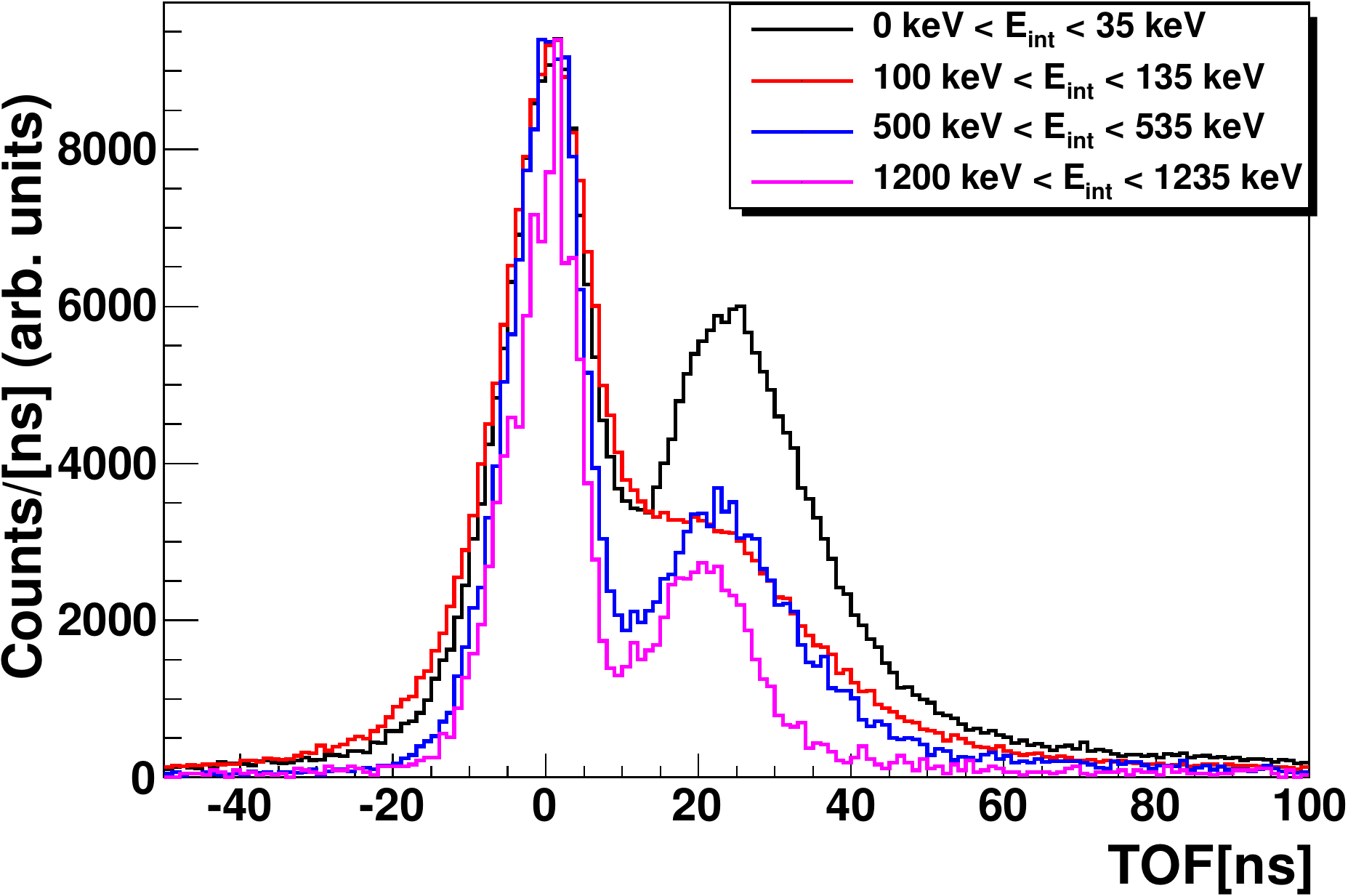}\\
  \caption{(Colour online) TOF histograms obtained with the
    \ce{^{252}Cf} source and with gates on different interaction point
    energy ranges.  The histograms are normalised to have the same
    height of the peak due to \g\ rays.}
  \label{f:fig5}
\end{figure}

The time resolution of the AGATA detectors can be investigated as a
function of interaction point energy by using the TOF histogram shown
in Fig.~\ref{f:fig3}a. Disregarding the time spent by \g\ rays
traversing the length of the Ge crystal ($\lesssim \SI{0.3}{\ns}$) and
the time resolution of the BaF$_{2}$ detectors (FWHM $\simeq
\SI{0.5}{\ns}$), the width of the time peak due to \g\ rays projected
on the y axis gives us an estimate of the time resolution of the AGATA
detectors. The results are shown in Fig.~\ref{f:fig5}, where TOF
histograms are plotted for different gates on the interaction point
energies. The time resolution is not so good for small interaction
point energies due to small detector signals with worse
signal-to-noise ratio. The \g\ rays give a time distribution with a
centroid at \SI{1.7}{\ns}, which is the time it takes for \g\ rays to
travel a distance of \SI{50}{\cm}. Since the flight time of the
neutrons depends on their energies, wider TOF distributions are
obtained for neutrons compared to \g\ rays.  For $E_{\textnormal{int}}
< \SI{35}{\keV}$, the enhanced neutron TOF distribution is mainly due
to recoiling Ge nuclei.  In the range of interaction point energies
from \SI{100}{\keV} to \SI{1200}{\keV}, the FWHM of the \g-ray time
peak decreases slowly from \SI{9.8}{\ns} to \SI{7.7}{\ns}.

The FWHM values given above for the AGATA detectors give a good
neutron-\g\ discrimination at a source to AGATA distance of
\SI{50}{\cm}. At the nominal distance of \SI{23.5}{\cm} the TOF method
can be useful only if a better time resolution is obtained, e.g. by
improving the PSA technique. This possibility was investigated by
Crespi et al.~\cite{Crespi2010299} by using a small non-segmented HPGe
detector.

\subsection{Tracking} \label{ss:ngd_tracking}

Neutron-\g\ discrimination based on standard tracking and on the three
tracking methods which were developed for improved neutron rejection,
are described in this subsection. These methods can be used instead
of, or as a complementary to, the TOF method.

\subsubsection{Standard tracking} \label{sss:standard_tracking}

As shown in Fig.~\ref{f:fig2}, the low-energy interaction points
($E_{\textnormal{int}} \lesssim \SI{45}{\keV}$), which are mainly due
the recoiling Ge nuclei after the elastic scattering of neutrons, are
effectively removed from the spectra by \g-ray tracking. In addition,
standard tracking also reduces the peaks and bumps due to inelastic
scattering of the neutrons in the HPGe detectors.

In order to obtain quantitative values for the reduction of counts in
the neutron-induced peaks and their associated bumps, the
\SI{1040}{\keV} peak, which originates from the
\ce{^{70}Ge}(n,n$^{\prime}\gamma$) reaction, was selected. In the
energy range of this peak and its bump (from \SI{1035}{\keV} to
\SI{1080}{\keV}) the \g-ray spectra are clean with no other peaks.  As
a reference, the reduction of counts in the \SI{1173}{\keV} and
\SI{1332}{\keV} peaks of \ce{^{60}Co} were also evaluated.  A perfect
neutron-\g\ discrimination would give a \SI{100}{\percent} reduction
of the counts both in the peak and in the bump due to the
\SI{1040}{\keV} transition, and no reduction (\SI{0}{\percent}) of the
counts in the \SI{1173}{\keV} and \SI{1332}{\keV} peaks.  The results
obtained with standard tracking were the following (see row~1 in
Table~\ref{t:tab4}): the \SI{1040}{\keV} peak was reduced by
\SI{57}{\percent}, the \SI{1040}{\keV} peak plus its associated bump
by \SI{52}{\percent}, while the \SI{1173}{\keV} and \SI{1332}{\keV}
peaks of \ce{^{60}Co} were reduced by \SI{9}{\percent} and
\SI{12}{\percent}, respectively.

\subsubsection{Tracking with neutron rejection: description of methods}
\label{sss:tracking_neutron_rejection_methods}

In order to improve the neutron rejection compared to what can be
achieved by using standard tracking, three methods based on the \g-ray
tracking were developed~\cite{Atac2009554}.  In this earlier work, the
interaction of neutrons and \g\ rays in the AGATA detectors were
simulated by using the \geant\ toolkit~\cite{Agostinelli2003250,
  Farnea2010331}.  Three methods were established to distinguish
between two types of \g\ rays: the ones that originate from the source
and the ones that are produced in the detectors after inelastic
scattering of neutrons, \ce{^{nat}Ge}(n,n$^{\prime}\gamma$).

The first method is based on the idea that the energy deposited in the
first interaction point ($E_{\textnormal{int},1}$) is expected to be
different for neutrons and \g\ rays. For neutron inelastic scattering,
the first interaction point, which is due to the Ge recoil, has a low
energy ($E_{\textnormal{int}} \lesssim \SI{35}{\keV}$, for
\SI{2}{\MeV} neutrons).  If the first interaction point in an \mgt\
cluster is identified correctly by the \g-ray tracking procedure, the
rejection of \mgt\ clusters with low $E_{\textnormal{int},1}$ values
can be a successful way of reducing neutron-induced
events. Considering the possibility that the ordering of the
interaction points in an \mgt\ cluster may be wrong, or that there may
be a neutron elastic scattering before the inelastic scattering, the
energy of the second interaction points ($E_{\textnormal{int},2}$) may
also be used for neutron rejection.

The second method is based on the difference in the incoming direction
of the \g\ rays, for which two different angles $\theta_{G}$ and
$\theta_{C}$, are defined~\cite{Ljungvall2005379, Recchia200960}.  The
geometric angle, $\theta_{G}$, is the angle between the line passing
through the position of the source and the first interaction point and
the line passing through the first and the second interaction points.
The Compton scattering angle, $\theta_{C}$, is calculated from the
Compton scattering formula using the position of the source, the total
energy deposited by the \g\ ray and the energy of the first
interaction point.  If the incoming particle is a \g\ ray, the
difference $\Delta\theta = \theta_{G} - \theta_{C}$ should be
distributed with its centroid at \SI{0}{\degree}. The width of the
distribution depends largely on the interaction position
resolution~\cite{Recchia200960, Recchia2009555} and on the ability of
the tracking algorithm to correctly assign the first and second
interaction points. If the incoming particle is a neutron, the
kinematics of the scattering will produce a different $\theta_{C}$
with respect to what is expected by the Compton scattering formula. A
broader asymmetric $\Delta\theta$ distribution is expected, with a
centroid that is not at \SI{0}{\degree}~\cite{Atac2009554}.  By
setting a gate on $\Delta\theta$ it may be possible to improve the
neutron rejection.

The third method is based on the selection of the FM value in the
tracking code. The validity of a cluster is checked by using its FM
value which is determined by the \g-ray interaction probabilities in
Ge. If a cluster contains a neutron interaction point, it is expected
to give a larger (worse) FM value compared to a cluster with only \g\
ray interaction points.

These three methods were tested in simulations of neutrons and \g\
rays emitted from the center of AGATA for different energy and
multiplicity values with good results for the neutron
rejection~\cite{Atac2009554}. In the present work, these tracking
methods were implemented in the \mgt\ program and tested with real
data.

\subsubsection{Tracking with neutron rejection: results}
\label{sss:tracking_neutron_rejection_results}

\begin{figure}[htbp]
  \centering
  \includegraphics[width=0.85\columnwidth]{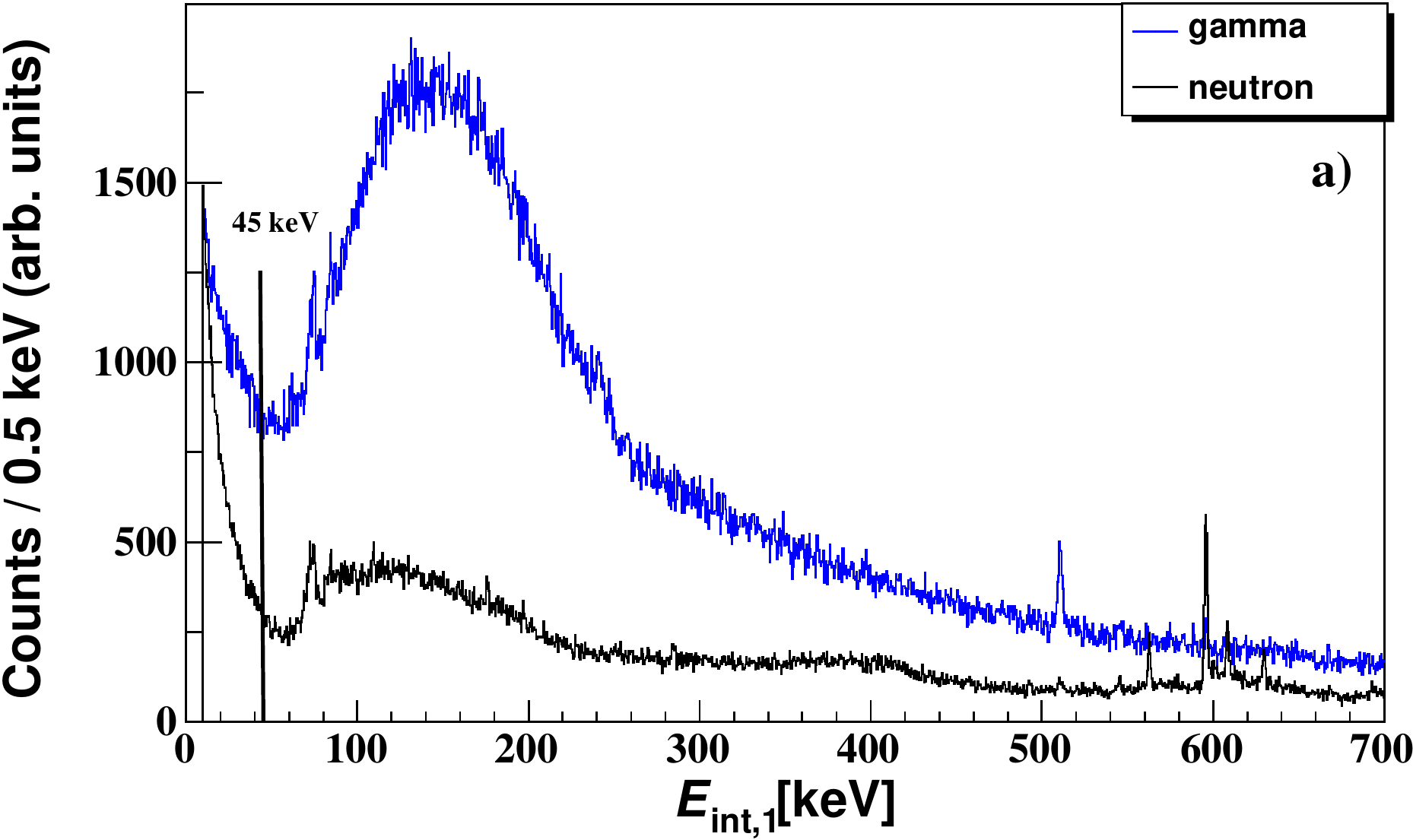}\\
  \includegraphics[width=0.85\columnwidth]{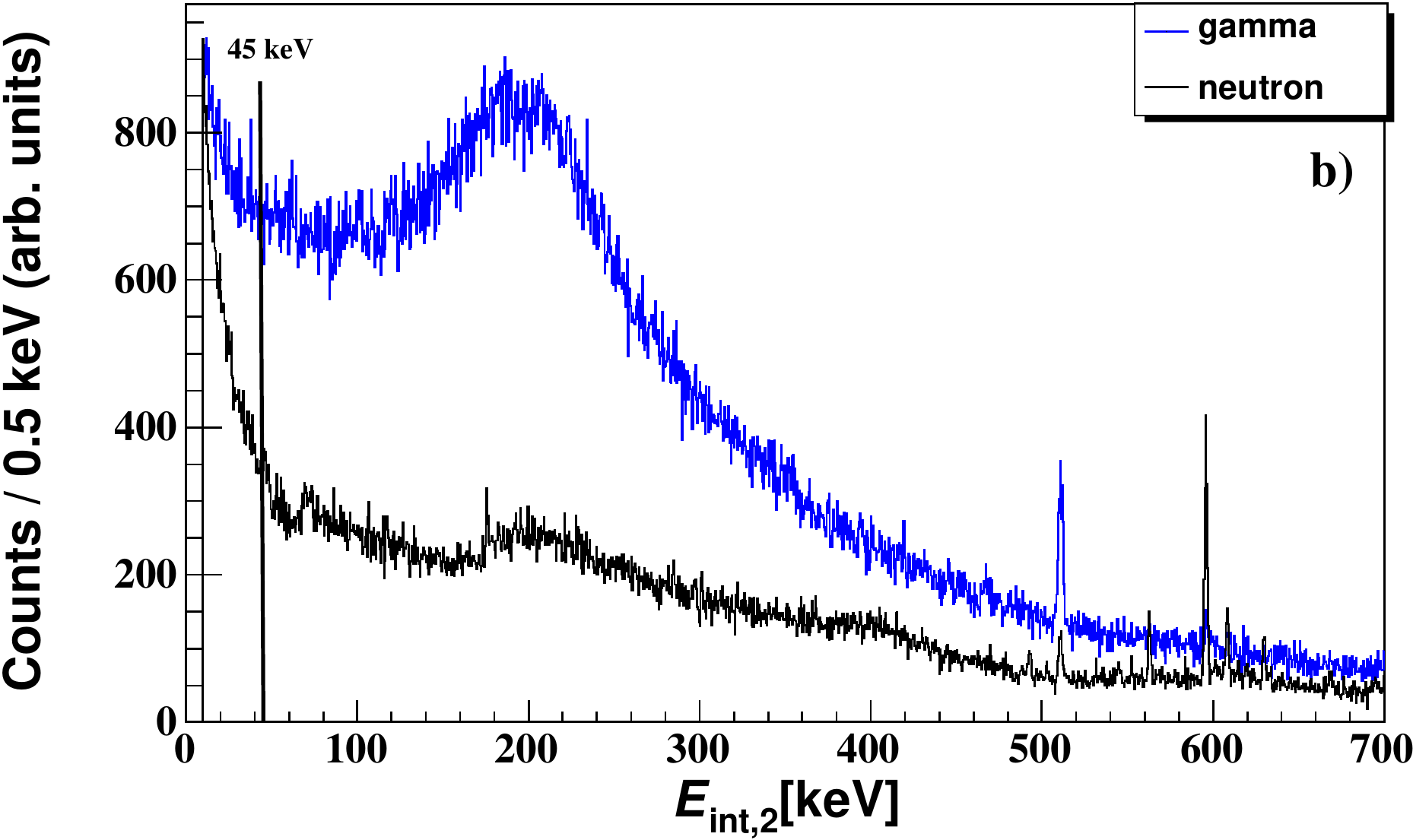}
  \caption{(Colour online) Distribution of the energy deposited in a)
    the first and b) the second interaction point by setting TOF gates
    on neutrons (black) and \g\ rays (blue) according to
    Fig.~\ref{f:fig3}a. The spectra were obtained with the
    \ce{^{252}Cf} source and by using standard tracking.  The spectra
    are normalised to have the same number of counts at \SI{10}{keV}.}
 \label{f:fig6}
\end{figure}

The spectra in Fig.~\ref{f:fig6} show the distribution of the energies
of the first ($E_{\textnormal{int},1}$) and second
($E_{\textnormal{int},2}$) interaction points for neutron and \g-ray
events, selected by using the n and g TOF gates, respectively, shown
in Fig.~\ref{f:fig3}a.  Both spectra, gated on neutrons, have large
abundances of counts at low energy, below about \SI{45}{\keV},
compared to the spectra gated on \g\ rays.  For example, by requiring
$E_{\textnormal{int},1} > \SI{45}{\keV}$ and/or
$E_{\textnormal{int},2} > \SI{45}{\keV}$, it is possible to eliminate
more of the neutron-induced events compared to the events that are due
to \g\ rays.

\begin{figure}[htbp]
  \centering
  \includegraphics[width=0.80\columnwidth]{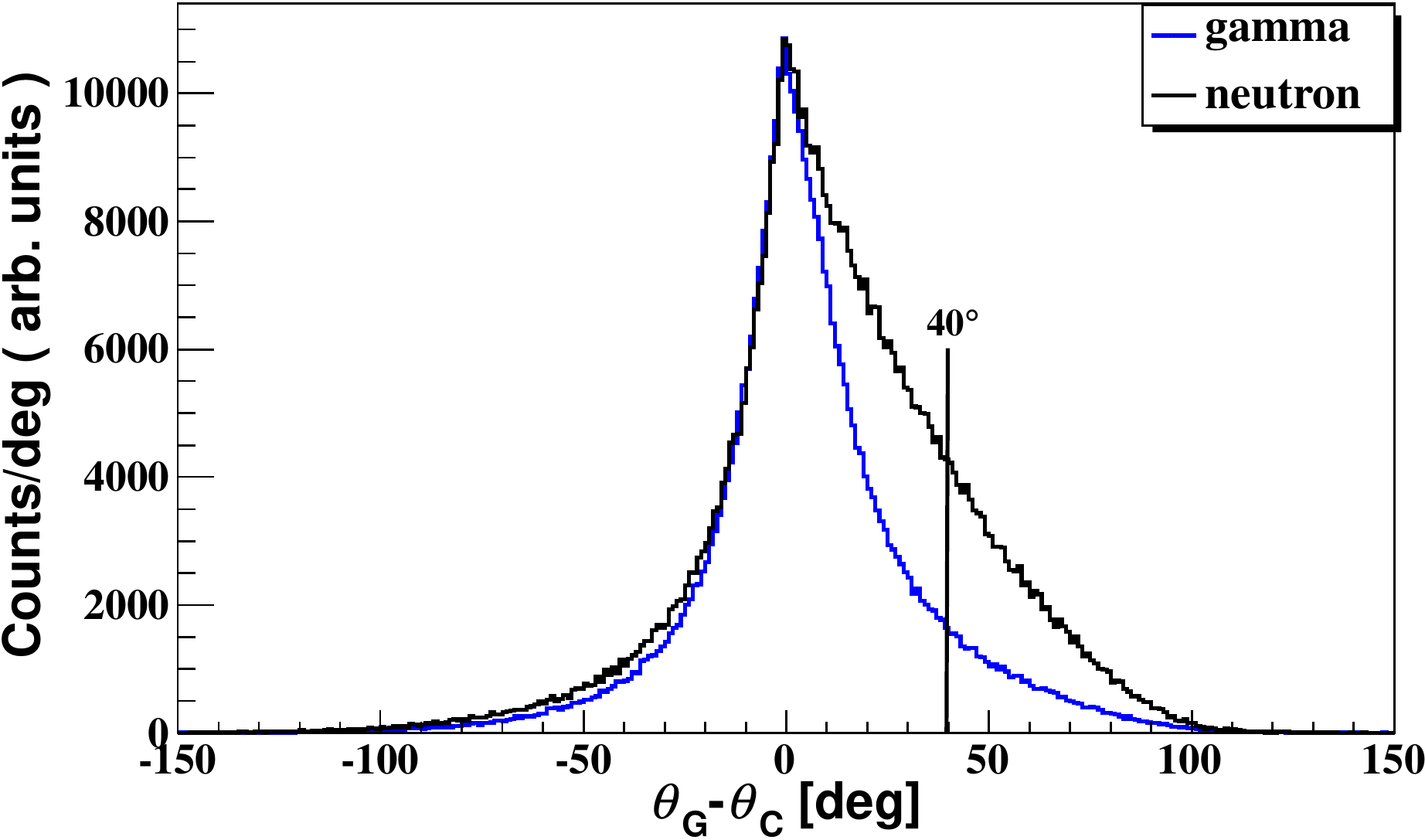}\\
  \caption{(Colour online) Distribution of $\Delta\theta =
    \theta_{G}-\theta_{C}$ for neutrons (black) and \g\ rays (blue)
    obtained with the \ce{^{252}Cf} source.  The neutrons and \g\ rays
    were selected by using the TOF gates n and g, respectively, shown
    in Fig.~\ref{f:fig3}a.  The spectra are normalised to have the
    same number of counts at the maximum of the peaks.}
  \label{f:fig7}
\end{figure}

The distribution of the difference
$\Delta\theta=\theta_{G}-\theta_{C}$ for neutrons and \g\ rays is
shown in Fig.~\ref{f:fig7}. As expected the distribution is centered
at \SI{0}{\degree} for \g\ rays, while a broader and asymmetric
distribution is observed for neutrons.  The two distributions are not
so different and a gate on $\Delta\theta$ for rejection of neutron
events, without loosing too many good \g-ray events, is not
trivial. Different values of the gate on $\Delta\theta$ were tested
and as a compromise $\Delta\theta < \SI{40}{\degree}$ was selected for
the further analysis below.

\begin{figure}[htbp]
  \centering
  \includegraphics[width=0.85\columnwidth]{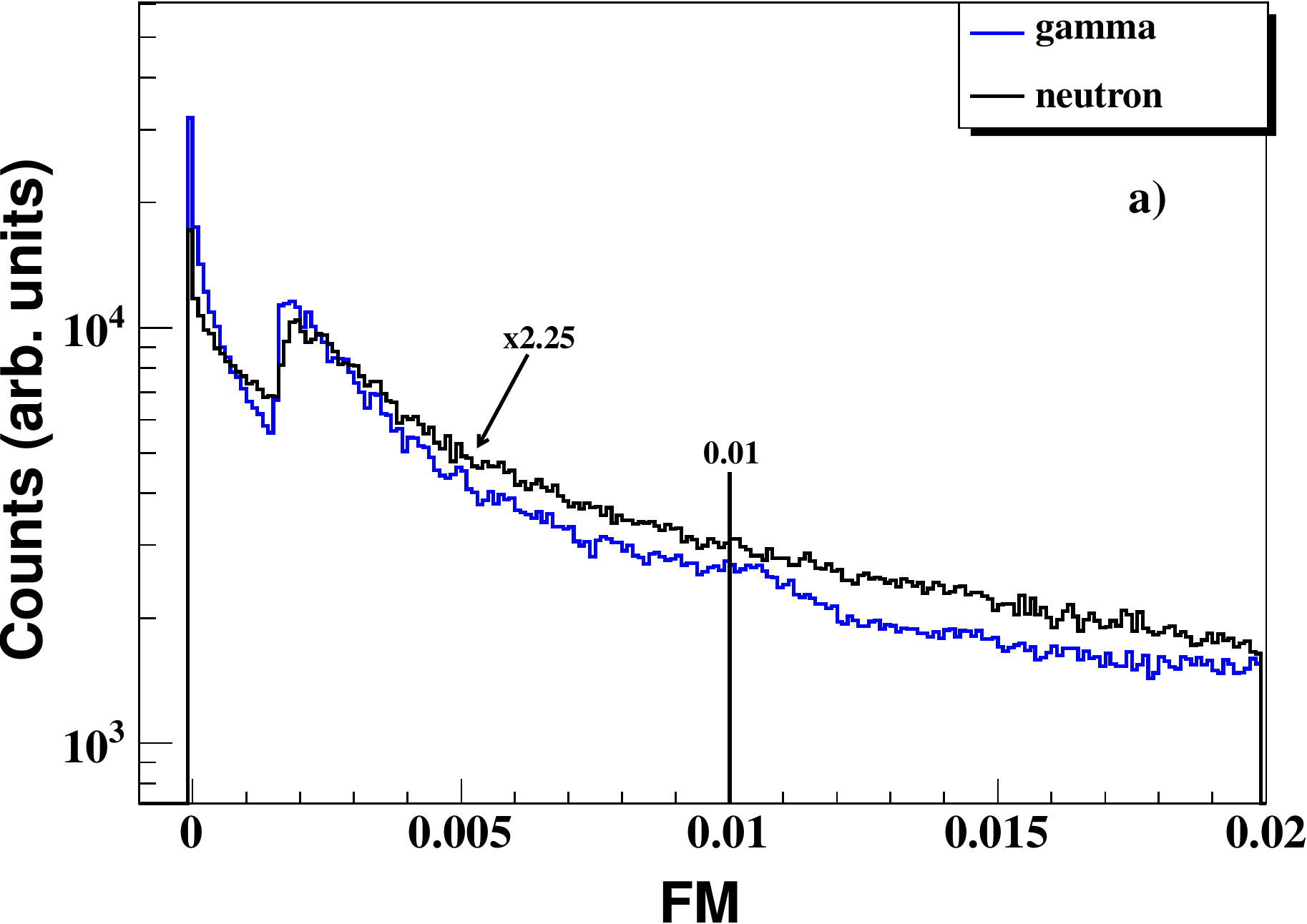}\\
  \includegraphics[width=0.85\columnwidth]{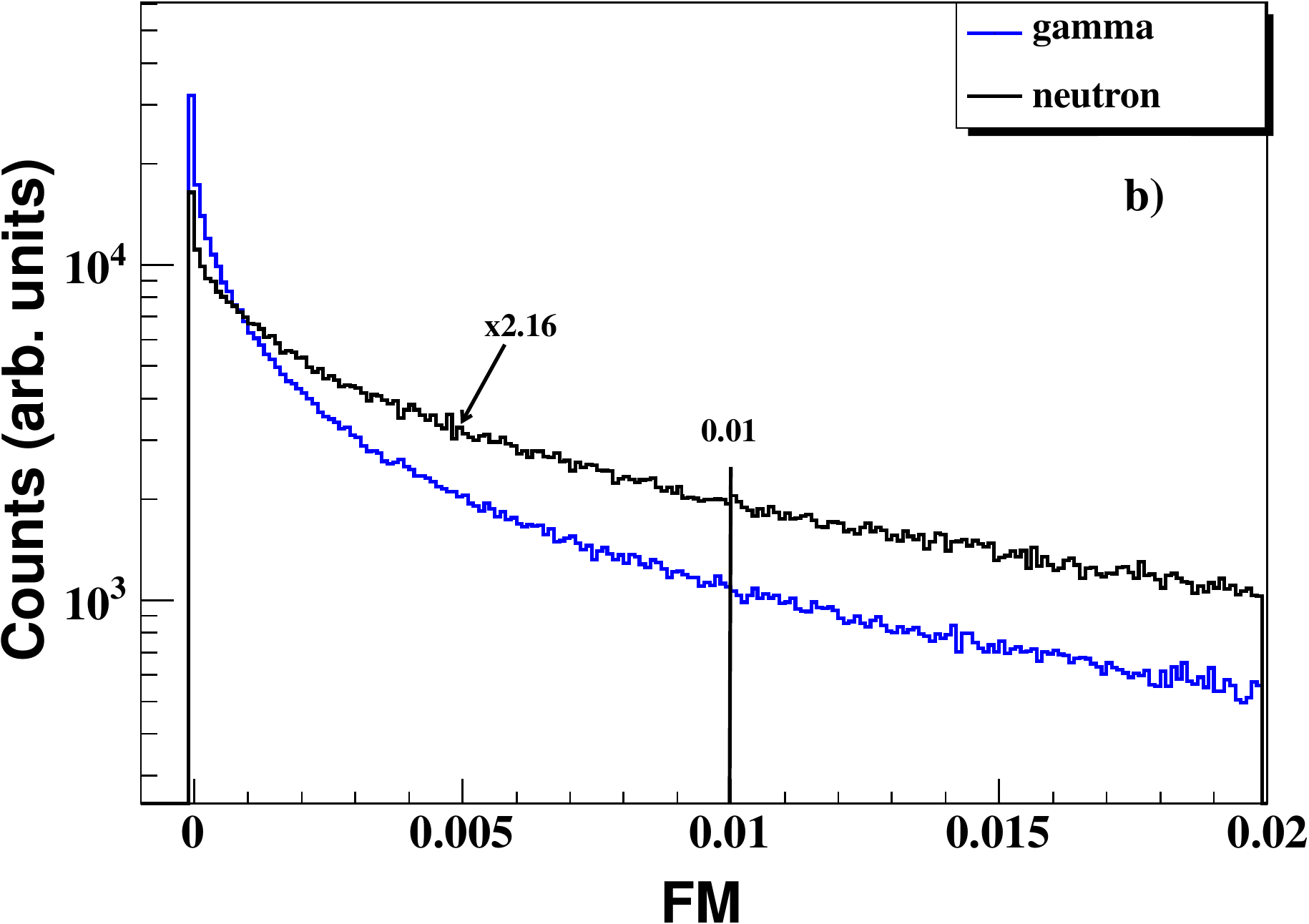}
  \caption{(Colour online) Distribution of the figure-of-merit (FM) of
    the \mgt\ clusters for data taken with the \ce{^{252}Cf} source
    for a) all events and b) events with $nptc>1$.  The blue spectrum
    is obtained by gating on the \g\ rays and black spectrum by gating
    on neutrons using the TOF gates show in Fig.~\ref{f:fig3}a. The
    blue and black histograms are normalised to have the same number
    of counts in the range of FM values from \num{0.4d-3} to
    \num{1.7d-3}.}
  \label{f:fig8}
\end{figure}

The distribution of FM values for neutrons and \g\ rays are shown in
Fig.~\ref{f:fig8} in the range from \num{0} to \num{0.02}. In this
work, the largest allowed FM value was \num{0.02}, which corresponds
to the \texttt{lim} value defined as standard tracking (see section
\ref{ss:eff_pt}).  There is almost no difference between the FM
distributions for neutrons and \g\ rays when events with any $nptc$
values are included (Fig.~\ref{f:fig8}a). The distributions for events
with $nptc>1$ (Fig.~\ref{f:fig8}b) show, however, that the neutron
induced events tend to have larger FM values.  Different FM values in
the range from \num{0} to \num{0.02} were tested, both for events with
$nptc>0$ and $nptc>1$, by analysing tracked \g-ray spectra and by
evaluating the rejection of neutron-induced events in the
\ce{^{252}Cf} dataset compared to rejection of \g-ray events in the
\ce{^{60}Co} dataset. The results of this evaluation led to the
conclusion that an FM value \num{<0.01} was optimal for the present
work. This is the value used in the further analysis below.

Different combinations of gates on $E_{\textnormal{int},1}$,
$E_{\textnormal{int},2}$, $\Delta\theta$ and FM were tested in order
to obtain the best neutron-\g\ discrimination. Quantitative results
are given in Table~\ref{t:tab2}.  The peak at \SI{1040}{\keV} ($2^{+}$
$\rightarrow0^{+}$ transition in \ce{^{70}Ge}) and its bump were
analysed in tracked spectra obtained from the \ce{^{252}Cf} dataset,
before and after different combinations of the four gates were
applied.  The \SI{1173}{\keV} and \SI{1332}{\keV} peaks of the
\ce{^{60}Co} dataset were analysed in the same way.  The aim was to
find the gate combinations, which reduce the number of counts in the
neutron-induced \SI{1040}{\keV} peak and its bump as much as possible,
while keeping the number of counts in the \ce{^{60}Co} peaks as large
as possible. The gate combinations that caused a loss of more than
\SI{20}{\percent} of the counts in the \SI{1332}{\keV} peak of
\ce{^{60}Co} were disregarded.

An example of a good gate combination is an OR of the following
conditions: $E_{\textnormal{int},1} > \SI{30}{\keV}$,
$E_{\textnormal{int},2} > \SI{30}{\keV}$, $\Delta\theta <
\SI{40}{\degree}$, $FM < 0.01$. This gate combination reduces the
\SI{1040}{\keV} transition by \SI{28}{\percent}, the \SI{1040}{\keV}
transition together with its bump by \SI{42}{\percent} and the total
spectrum from \SI{0}{\keV} to \SI{4095}{\keV} by \SI{34}{\percent},
compared to the spectrum obtained by standard tracking.  This caused a
loss of \SI{19}{\percent} of the counts in the \SI{1332}{\keV} peak of
\ce{^{60}Co}. An investigation of correlations between the gating
parameters, which could be used to further improve the neutron
rejection, was also made.  This was done by producing two-dimensional
histograms of the parameters, e.g.  $E_{\textnormal{int},1}$ versus
$\Delta\theta$ or $E_{\textnormal{int},1}$ versus FM, and by setting
two-dimensional gates in such histograms. No apparent improvement of
the neutron rejection was obtained in this way.

\begin{table*}[htbp]
  \centering
  \caption {Reduction of the number of counts (in percent) in 
    experimental tracked \g-ray spectra when applying different
    combinations of the neutron-\g\ discrimination gates. Errors
    given are purely statistical. See text for details.}
  \begin{tabular}{|c|c|c|c|c|c|}
    \hline
    \multirow{3}{*}{Gate} &
    \multicolumn{3}{c|}{\ce{^{252}Cf}} & 
    \multicolumn{2}{c|}{\ce{^{60}Co}} \\
    \cline{2-6}
    & \multicolumn{2}{c|}{\SI{1040}{\keV}} & Total & 
    \multirow{2}{*}{\SI{1173}{\keV}} & \multirow{2}{*}{\SI{1332}{\keV}} \\
    \cline{2-3}
    & peak & peak $+$ bump & \num{0}-\SI{4095}{\keV} & & \\
    \hline
    $E_{\textnormal{int},1} > \SI{45}{\keV}$ & 
    $2(3)$  & $16(3)$ & $7.2(1)$  & $4.7(6)$ & $3.9(6)$ \\
    \hline
    $E_{\textnormal{int},2} > \SI{45}{\keV}$ & 
    $4(3)$  & $9(3)$ & $5.7(1)$  & $5.1(7)$ & $4.9(6)$ \\
    \hline
    $\Delta\theta < \SI{40}{\degree}$ & 
    $9(3)$ & $13(3)$ & $6.8(1)$  & $6.7(6)$ & $7.2(6)$ \\
    \hline
    $FM < 0.01$ & 
    $18(3)$ & $23(3)$ & $25.3(1)$ & $9.8(6)$ & $10.2(6)$ \\
    \hline
    $E_{\textnormal{int},1} > \SI{30}{\keV}$ OR &
    \multirow{2}{*}{$10(2)$}    & \multirow{2}{*}{$21(3)$}  & 
    \multirow{2}{*}{$10.4(1)$} & \multirow{2}{*}{$8.6(6)$} &
    \multirow{2}{*}{$8.9(6)$} \\
    $\Delta\theta < \SI{40}{\degree}$     &       &       &       &     & \\
    \hline
    $E_{\textnormal{int},2} > \SI{30}{\keV}$ OR &
    \multirow{2}{*}{$11(3)$}    & \multirow{2}{*}{$20(3)$}  & 
    \multirow{2}{*}{$9.9(1)$} & \multirow{2}{*}{$9.6(6)$} &
    \multirow{2}{*}{$9.9(6)$} \\
    $\Delta\theta < \SI{40}{\degree}$     &       &       &       &     & \\
    \hline
    $E_{\textnormal{int},1} > \SI{40}{\keV}$ OR &
    \multirow{3}{*}{$28(3)$}    & \multirow{3}{*}{$42(2)$} &
    \multirow{3}{*}{$35.4(1)$} & \multirow{3}{*}{$19.9(6)$} &
    \multirow{3}{*}{$20.1(6)$} \\
    $E_{\textnormal{int},2} > \SI{40}{\keV}$ OR &       &       &       &    & \\
    $\Delta\theta < \SI{40}{\degree}$ OR $FM < 0.01$ &  &  &       &    & \\
    \hline
    $E_{\textnormal{int},1} > \SI{30}{\keV}$ OR &
    \multirow{3}{*}{$28(3)$}    & \multirow{3}{*}{$42(3)$} &
    \multirow{3}{*}{$33.7(1)$} & \multirow{3}{*}{$18.0(6)$} &
    \multirow{3}{*}{$18.5(6)$} \\
    $E_{\textnormal{int},2} > \SI{30}{\keV}$ OR &       &       &        &    & \\
    $\Delta\theta < \SI{40}{\degree}$ OR $FM < 0.01$ &  &          &    & \\
    \hline
  \end{tabular}
  \label{t:tab2}
\end{table*}

In order to compare the experimental results with simulations, a
simulation was carried out with the AGATA \geant\
code~\cite{Farnea2010331} using a setup with four ATC detectors placed
at a distance of \SI{50}{\cm} from the \ce{^{252}Cf} and \ce{^{60}Co}
sources.  The simulation was performed as described in detail in
Ref.~\cite{Atac2009554}, except that the number of ATCs and the
distance to the source was different.  Tracking was performed by using
$lim=0.02$ and two different low-energy thresholds, \SI{5}{\keV} and
\SI{10}{\keV}.  The results of the simulation are given in
Table~\ref{t:tab3}. By using a low-energy threshold of \SI{10}{\keV}
and the gate combination [$E_{\textnormal{int},1} > \SI{30}{\keV}$ OR
$E_{\textnormal{int},1} > \SI{30}{\keV}$ OR $\Delta\theta <
\SI{40}{\degree}$ OR $FM < 0.01$], the counts in the \SI{1040}{\keV}
peak was reduced by \SI{25}{\percent}, in the \SI{1040}{\keV} peak and
its associated bump by \SI{39}{\percent} and in the total spectrum
(\SI{0}{\keV} to \SI{4095}{\keV}) by \SI{43}{\percent}, as compared to
the spectrum obtained by standard tracking.  This caused a loss of
\SI{12}{\percent} of the counts in the \SI{1332}{\keV} peak.

The effect of varying the low-energy threshold was tested by reducing
the value to \SI{5}{\keV}, which gave a significant improvement of the
neutron rejection. In this case, with the same gate combination and
having nearly the same loss of counts in the \SI{1332}{\keV} peak, the
counts in the \SI{1040}{\keV} peak was reduced by \SI{28}{\percent},
in the peak plus the bump by \SI{57}{\percent} and in the total
spectrum by \SI{51}{\percent}, as compared to the spectrum obtained by
standard tracking.

The experimental results are in rather good agreement with the results
of the simulation obtained by using a low-energy threshold of
\SI{10}{\keV} (compare the results in Table~\ref{t:tab2} and
\ref{t:tab3}). The experimental results are slightly worse than the
simulated results, in particular regarding the reduction of the counts
in the \SI{1332}{\keV} peak of \ce{^{60}Co}.  It is likely that the
experimental results can become better by improving the PSA techniques
to give more precise interaction positions and better identification
of multiple hits in the same crystal.  Note that, with such an
improvement, it may be possible to use only one or two of the gates,
e.g. the combination [$E_{\textnormal{int},1} > \SI{30}{\keV}$ OR
$\Delta\theta < \SI{40}{\degree}$], to obtain a reasonable reduction
of the neutron-induced background without loosing efficiency in the
\ce{^{60}Co} peaks.

\begin{table*}[htbp]
  \centering
  \caption {Reduction of the number of counts (in percent) in \geant\ 
    simulated tracked \g-ray spectra when applying different combinations
    of the neutron-\g\ discrimination gates and by using two different 
    low-energy thresholds, \SI{5}{\keV} and \SI{10}{\keV}, on the 
    interaction point energies. Errors given are purely statistical.}
  \begin{tabular}{|c|c|c|c|c|c|}
    \hline
    \multirow{3}{*}{} & \multirow{3}{*}{Gate} &
    \multicolumn{3}{c|}{\ce{^{252}Cf}} & 
    \multirow{3}{*}{\SI{1332}{\keV}} \\
    \cline{3-5}
    & & \multicolumn{2}{c|}{\SI{1040}{\keV}} & Total & \\
    \cline{3-4}
    & & peak & peak $+$ bump & \num{0}-\SI{4095}{\keV}  &  \\
    \hline
    \multirow{6}{*}{\rotatebox{90}
      {\mbox{$E_{\textnormal{int, thr}} = \SI{5}{\keV}$}}}
    & & & & & \\
    & $E_{\textnormal{int},1} > \SI{45}{\keV}$ &
    $2(5)$ & $20(4)$ & $15.2(4)$ & $2(2)$ \\
    \cline{2-6}
    & $E_{\textnormal{int},2} > \SI{45}{\keV}$ & 
    $12(5)$  & $30(4)$ & $22.5(4)$ & $8(3)$ \\
    \cline{2-6}
    & $\Delta\theta < \SI{40}{\degree}$ & 
    $6(5)$ & $13(4)$ & $15.4(4)$ & $3(2)$ \\
    \cline{2-6}
    & $FM < 0.01$ & 
    $16(5)$ & $26(4)$ & $25.7(4)$ & $5(2)$ \\
    \cline{2-6}
    & $E_{\textnormal{int},1} > \SI{30}{\keV}$ OR
    $\Delta\theta < \SI{40}{\degree}$ & 
    $8(5)$ & $26(4)$ & $25.5(4)$ & $4(2)$ \\
    \cline{2-6}
    & $E_{\textnormal{int},2} > \SI{30}{\keV}$ OR
    $\Delta\theta < \SI{40}{\degree}$ & 
    $14(5)$ & $38(4)$ & $31.1(4)$ & $5(3)$ \\
    \cline{2-6}
    & $E_{\textnormal{int},1} > \SI{30}{\keV}$ OR 
    $E_{\textnormal{int},2} > \SI{30}{\keV}$ &
    \multirow{2}{*}{$28(5)$}   & \multirow{2}{*}{$57(4)$} &
    \multirow{2}{*}{$51.3(4)$} & \multirow{2}{*}{$13(2)$} \\
    & OR $\Delta\theta < \SI{40}{\degree}$ OR $FM < 0.01$ & &  &  & \\
    \hline
    \multirow{6}{*}{\rotatebox{90}
      {\mbox{$E_{\textnormal{int, thr}} = \SI{10}{\keV}$}}}
    & & & & & \\
    & $E_{\textnormal{int},1} > \SI{45}{\keV}$ & 
    $1(4)$ & $10(3)$ & $9.2(4)$ & $2(4)$ \\
    \cline{2-6}
    & $E_{\textnormal{int},2} > \SI{45}{\keV}$ & 
    $10(4)$  & $17(3)$ & $16.0(4)$  & $7(4)$ \\
    \cline{2-6}
    & $\Delta\theta < \SI{40}{\degree}$  &
    $7(4)$  & $10(3)$ & $14.5(4)$ & $3(4)$ \\
    \cline{2-6}
    & $FM < 0.01$                        & 
    $14(4)$ & $18(3)$ & $22.3(4)$ & $5(4)$ \\
    \cline{2-6}
    & $E_{\textnormal{int},1} > \SI{30}{\keV}$ OR 
    $\Delta\theta < \SI{40}{\degree}$ & 
    $7(4)$ & $15(3)$ & $20.2(4)$ & $4(4)$ \\
    \cline{2-6}
    & $E_{\textnormal{int},2} > \SI{30}{\keV}$ OR 
    $\Delta\theta < \SI{40}{\degree}$ & 
    $12(4)$ & $22(3)$ & $24.0(4)$ & $5(4)$ \\
    \cline{2-6}
    & $E_{\textnormal{int},1} > \SI{30}{\keV}$ OR
    $E_{\textnormal{int},2} > \SI{30}{\keV}$ & 
    \multirow{2}{*}{$25(4)$} & \multirow{2}{*}{$39(3)$} & 
    \multirow{2}{*}{$42.5(4)$} & \multirow{2}{*}{$12(4)$} \\
    & OR $\Delta\theta < \SI{40}{\degree}$ OR $FM < 0.01$ & &  &  & \\
    \hline
  \end{tabular}
  \label{t:tab3}
\end{table*}

\begin{figure*}[htbp]
  \centering
  \includegraphics[width=0.99\textwidth]{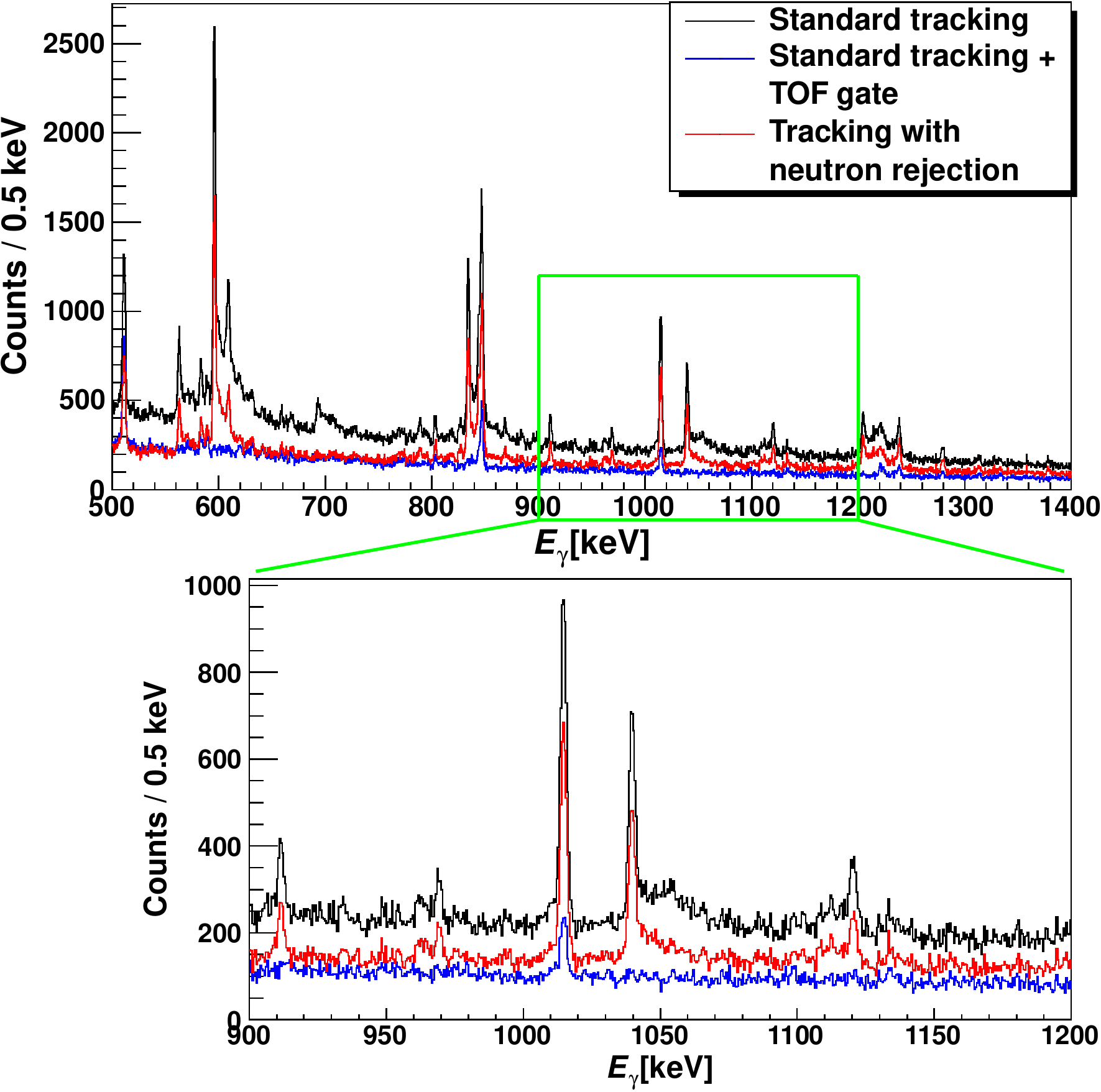}\\
  \caption{(Colour online) Tracked \g-ray energy spectra obtained with
    the \ce{^{252}Cf} source and by using standard tracking (black),
    standard tracking $+$ TOF gate on \g\ rays (blue), and tracking
    with neutron rejection using the gate combination
    [$E_{\textnormal{int},1} > \SI{30}{\keV}$ OR
    $E_{\textnormal{int},2} > \SI{30}{\keV}$ OR $\Delta\theta <
    \SI{40}{\degree}$ OR $FM < 0.01$] (red).  The spectra are shown in
    the energy range were most of the peaks due to inelastic
    scattering of neutrons in the HPGe crystals are located.  An
    expanded version of the region around the \SI{1040}{\keV} peak
    ($2^{+}$ $\rightarrow0^{+}$ transition in \ce{^{70}Ge}) is also
    shown.}
  \label{f:fig9}
\end{figure*}

Since the TOF method gives an excellent neutron rejection, the \g-ray
spectrum obtained by using the TOF gate is compared to the one which
is obtained with the gate combination [$E_{\textnormal{int},1} >
\SI{30}{\keV}$ OR $E_{\textnormal{int},2} > \SI{30}{\keV}$ OR
$\Delta\theta < \SI{40}{\degree}$ OR $FM < 0.01$], in
Fig.~\ref{f:fig9}. Apart from the reduction of the neutron-induced
peaks. e.g. the peaks at \SI{596}{\keV}, \SI{834}{\keV},
\SI{1040}{\keV}, and \SI{1204}{\keV}, and in their associated bumps,
the overall background is also reduced successfully in the spectrum
produced by this gate combination.

\begin{figure*}[htbp]
  \centering
  \includegraphics[width=0.99\textwidth]{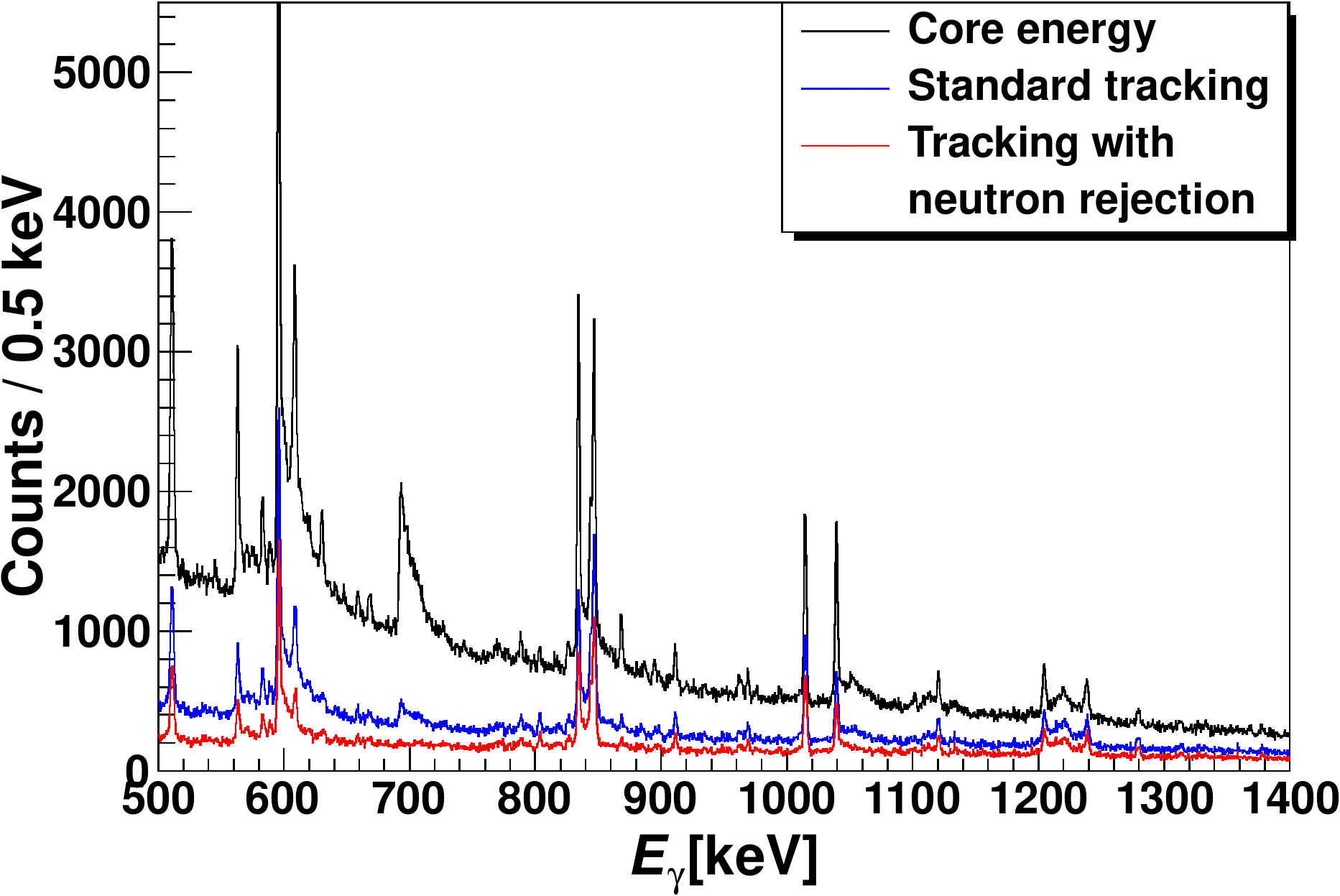}\\
  \caption{(Colour online) Gamma-ray energy spectra measured with the
    \ce{^{252}Cf} source and displayed in the energy range were most
    of the peaks due to inelastic scattering of neutrons in the HPGe
    crystals are located.  The following spectra are shown:
    core-energy spectrum (black), tracked spectrum obtained by
    standard tracking (blue) and tracked spectrum with neutron
    rejection (red), using the gate combination
    [$E_{\textnormal{int},1} > \SI{30}{\keV}$ OR
    $E_{\textnormal{int},2} > \SI{30}{\keV}$ OR $\Delta\theta <
    \SI{40}{\degree}$ OR $FM < 0.01$].  }
  \label{f:fig10}
\end{figure*}

An interesting question is how much the neutron-induced background is
removed in the \g-ray spectra produced by standard tracking and by
tracking with neutron rejection, compared to the core-energy spectra.
A comparison between these three spectra is shown in
Fig.~\ref{f:fig10} and quantitative results are given in
Table~\ref{t:tab4}. As mentioned in
section~\ref{sss:standard_tracking}, a spectrum created by standard
tracking already rejects a considerable amount of the neutron-induced
events compared to a core-energy spectrum (see row 1 in
Table~\ref{t:tab4}).  By using tracking with neutron rejection the
spectra are further improved (row 3 in Table~\ref{t:tab4}): the number
of counts in the \SI{1040}{\keV} peak is reduced by \SI{69}{\percent},
its bump by \SI{72}{\percent} and the total \ce{^{252}Cf} \g-ray
spectrum by \SI{83}{\percent} compared to the core-energy
spectrum. This loss of counts in the \SI{1332}{\keV} peak of
\ce{^{60}Co} was in this case \SI{19}{\percent} compared to the
spectrum obtained by standard tracking and \SI{26}{\percent} compared
to the core-energy spectrum.

\begin{table*}[htbp]
  \centering
  \caption{Reduction of the number of counts (in percent) in a spectrum
    obtained by standard tracking and in a spectrum obtained by using
    tracking with neutron rejection compared to a core-energy spectrum.
    The gate used for the spectrum made by tracking with neutron rejection was  
    [$E_{\textnormal{int},1} > \SI{30}{\keV}$ OR 
    $E_{\textnormal{int},2} > \SI{30}{\keV}$ OR
    $\Delta\theta < \SI{40}{\degree}$ OR $FM < 0.01$].
    Results are shown for the \SI{1040}{\keV} peak, for 
    \SI{1040}{\keV} peak plus its associated bump and for the total 
    spectrum (\num{0}-\SI{4095}{\keV}) obtained with the \ce{^{252}Cf} 
    source, as well as for the \SI{1173}{\keV} and  \SI{1332}{\keV} 
    obtained with the \ce{^{60}Co} source.}
  \begin{tabular}{|c|c|c|c|c|c|}
    \hline
    \multirow{3}{*}{Compared spectra} &
    \multicolumn{3}{c|}{\ce{^{252}Cf}} & 
    \multicolumn{2}{c|}{\ce{^{60}Co}} \\
    \cline{2-6}
    & \multicolumn{2}{c|}{\SI{1040}{\keV}} & Total &
    \multirow{2}{*}{\SI{1173}{\keV}} &
    \multirow{2}{*}{\SI{1332}{\keV}} \\
    \cline{2-3}
    &  peak & peak $+$ bump & \num{0}-\SI{4095}{\keV} & & \\
    \hline
    core energy vs &
    \multirow{2}{*}{$57(1)$}    & \multirow{2}{*}{$52(1)$}   &
    \multirow{2}{*}{$73.73(2)$} & \multirow{2}{*}{$11.7(6)$} &
    \multirow{2}{*}{$9.0(6)$} \\
    standard tracking &
    & & & & \\
    \hline
    standard tracking &
    \multirow{3}{*}{$28(2)$}    & \multirow{3}{*}{$42(2)$}   &
    \multirow{3}{*}{$33.68(8)$} & \multirow{3}{*}{$18.0(6)$} &
    \multirow{3}{*}{$18.5(6)$} \\
    vs tracking with &
    & & & & \\
    neutron rejection &
    & & & & \\
    \hline
    core energy vs &
    \multirow{3}{*}{$69(1)$}    & \multirow{3}{*}{$72(1)$}   &
    \multirow{3}{*}{$82.58(2)$} & \multirow{3}{*}{$27.6(5)$} &
    \multirow{3}{*}{$25.8(5)$} \\
    tracking with &
    & & & & \\
    neutron rejection &
    & & & & \\
    \hline
  \end{tabular}
  \label{t:tab4}
\end{table*}

%%-----------------------------------------------------------------------------

\section{Summary and conclusions} \label{s:conclusions}

In this work, \g\ rays originating from inelastic scattering of
neutrons from a \ce{^{252}Cf} source, which was placed at a distance
of \SI{50}{\cm} from four AGATA Triple Cluster detectors, were
identified and rejected from tracked \g-ray energy spectra.  Several
methods for neutron rejection, based on \g-ray tracking and developed
in a \geant\ simulation work~\cite{Atac2009554}, were tested with
experimental data.  The methods were optimised with a special emphasis
on not reducing the full-energy peak efficiency of the \g\ rays of
interest.

A time-of-flight measurement, using \num{16} HELENA BaF$_{2}$
detectors as time reference, was performed. This was done in order to
have an independent and well understood discrimination of neutrons and
\g\ rays as well as to study the possibility of using TOF
discrimination with AGATA.  We have shown that the TOF discrimination
of neutrons and \g\ rays, emitted by a \ce{^{252}Cf} source, works
very well for an ATC to source distance of \SI{50}{\cm} or
larger. However, at the nominal ATC to source distance of
\SI{23.5}{\cm} or at shorter distances, the TOF discrimination will be
of limited use with the presently achievable time resolution of the
HPGe detectors.  In such cases, the tracking methods for neutron
rejection, developed in the previous work~\cite{Atac2009554} and
tested here, may be important tools for neutron-\g\ discrimination.

The standard \g-ray tracking method eliminates effectively most of the
interaction points that are due to the recoiling Ge nuclei after
elastic scattering of neutrons. Apart from this, standard tracking
reduces also the background due to inelastic scattering of neutrons.
For example, the \SI{1040}{\keV} peak, which originates from the
\ce{^{70}Ge}(n,n$^{\prime}\gamma$) reaction, was reduced by
\SI{57}{\percent}, the counts in this peak and its associated bump,
due to the summing of the Ge-recoil energies, by \SI{52}{\percent} and
the total spectrum from \SI{0}{\keV} to \SI{4095}{\keV} by
\SI{74}{\percent}, as compared to the a \g-ray energy spectrum created
from the individual HPGe core signals.  In the measurement with a
\ce{^{60}Co} source, standard tracking caused a loss of
\SI{9}{\percent} of the counts in the \SI{1332}{\keV} peak compared to
the core-energy spectrum.

Further discrimination of neutron-induced events was achieved by using
the tracking-based neutron rejection methods.  With the combined
condition [$E_{\textnormal{int},1} < \SI{30}{\keV}$ OR
$E_{\textnormal{int},2} < \SI{30}{\keV}$ OR $\Delta\theta >
\SI{40}{\degree}$ OR $FM > 0.01$] on the first and second interaction
point energies ($E_{\textnormal{int},1}$, $E_{\textnormal{int},2}$),
on the difference in the evaluated incoming direction of the \g\ ray
($\Delta\theta$), or on the figure-of-merit value of the tracking
algorithm, the following results were obtained: the \SI{1040}{\keV}
transition was reduced by \SI{28}{\percent}, the \SI{1040}{\keV}
transition together with its bump by \SI{42}{\percent} and the total
spectrum from \SI{0}{\keV} to \SI{4095}{\keV} by \SI{34}{\percent},
compared to a spectrum obtained by standard tracking.  This caused a
loss of \SI{19}{\percent} of the counts in the \SI{1332}{\keV} peak
compared to standard tracking.

The experimental results for tracking with neutron rejection are in
reasonably good agreement with the \geant\ AGATA simulations.  It is
expected that the results will be improved even further by reducing
the low-energy thresholds of the ATC detectors and by improving the
pulse-shape analysis techniques for better interaction position
resolution.

%%-----------------------------------------------------------------------------

\section{Acknowledgments } \label{s:acknowledgments}

This work was partly supported by the Scientific and Technological
Council of Turkey (project number 106T055) and by the Swedish Research
Council.

%%-----------------------------------------------------------------------------

% References with bibTeX database:

\bibliographystyle{./Bibliography/jn-nima}
\bibliography{./Bibliography/references}

\begin{thebibliography}{10}
\providecommand{\url}[1]{\texttt{#1}}
\providecommand{\urlprefix}{}

\bibitem{Akkoyun201226}
S.~Akkoyun et~al., Nucl. Instr. Meth. A 668 (2012) 26--58.

\bibitem{Deleplanque1999292}
M.~A. Deleplanque et~al., Nucl. Instr. Meth. A 430 (1999) 292 -- 310.

\bibitem{Lee2003}
I.~Y. Lee, M.~A. Deleplanque, and K.~Vetter, Rep. Prog. Phys. 66 (2003) 1095.

\bibitem{Lee2004}
I.~Y. Lee et~al., Nucl. Phys. A {746} ({2004}) {255--259}.

\bibitem{Agostinelli2003250}
S.~Agostinelli et~al., Nucl. Instr. Meth. A 506 (2003) 250 -- 303.

\bibitem{Farnea2010331}
E.~Farnea et~al., Nucl. Instr. Meth. A 621 (2010) 331 -- 343.

\bibitem{Ljungvall2005379}
J.~Ljungvall and J.~Nyberg, Nucl. Instr. Meth. A 550 (2005) 379 -- 391.

\bibitem{Atac2009554}
A.~Ata\c{c} et~al., Nucl. Instr. Meth. A 607 (2009) 554 -- 563.

\bibitem{Ljungvall2005553}
J.~Ljungvall and J.~Nyberg, Nucl. Instr. Meth. A 546 (2005) 553 -- 573.

\bibitem{Jenkins2009457}
D.~G. Jenkins et~al., Nucl. Instr. Meth. A 602 (2009) 457 -- 460.

\bibitem{Abt2008}
I.~Abt et~al., Eur. Phys. J. A 36 (2008) 139--149.

\bibitem{Gadea201188}
A.~Gadea et~al., Nucl. Instr. Meth. A 654 (2011) 88 -- 96.

\bibitem{Farnea2012}
E.~Farnea, AIP Conf. Proc. 1491 (2012) 42.

\bibitem{Wiens2010223}
A.~Wiens et~al., Nucl. Instr. Meth. A 618 (2010) 223 -- 233.

\bibitem{Boiano2008}
C.~Boiano et~al., IEEE Nucl. Sci. Symp. Conf. Rec.  (2008) N30--46.

\bibitem{Venturelli2005}
R.~Venturelli and D.~Bazzacco, {INFN-LNL Annual Report 2004}, INFN-LNL (2005)
  220.

\bibitem{adl}
B.~Bruyneel, {AGATA Detector Simulation Software ADL},
  \urlprefix\url{http://www.ikp.uni-koeln.de/research/agata/download.php},
  unpublished.

\bibitem{Bazzacco2004248}
D.~Bazzacco, Nucl. Phys. A 746 (2004) 248 -- 254.

\bibitem{Schmid199969}
G.~J. Schmid et~al., Nucl. Instr. Meth. A 430 (1999) 69 -- 83.

\bibitem{LopezMartens2004454}
A.~Lopez-Martens et~al., Nucl. Instr. Meth. A 533 (2004) 454 -- 466.

\bibitem{radware}
D.~C. Radford, {\texttt{GF3} program},
  \urlprefix\url{http://radware.phy.ornl.gov/}, unpublished.

\bibitem{root}
R.~Brun and F.~Rademakers, Nucl. Instr. Meth. A 389 (1997) 81 -- 86.

\bibitem{Knoll2010}
G.~F. Knoll, {Radiation Detection and Measurement}, John Wiley \& Sons, Inc.,
  fourth edition edition (2010), and references therein.

\bibitem{Valentine2001191}
T.~E. Valentine, Ann. Nucl. Energy 28 (2001) 191 -- 201, and references
  therein.

\bibitem{Bleuel2010691}
D.~L. Bleuel et~al., Nucl. Instr. Meth. A 624 (2010) 691 -- 698.

\bibitem{Ensslin1991}
N.~Ensslin, {Passive Nondestructive Assay of Nuclear Materials}, Los Alamos
  National Laboratory (1991) 337 -- 356,
  \urlprefix\url{http://www.fas.org/sgp/othergov/doe/lanl/lib-www/la-pubs/00326406.pdf}.

\bibitem{Shen2002}
Q.~B. Shen, {CENDL-3.1 MAT 3200} (September 2002).

\bibitem{Iwamoto2004}
O.~Iwamoto et~al., {ENDF/B-VII MAT 3225} (August 2004).

\bibitem{ShamsuzzohaBasunia20111875}
M.~S. Basunia, Nucl. Data Sheets 112 (2011) 1875 -- 1948.

\bibitem{Elenkov1982}
D.~V. Elenkov, D.~P. Lefterov, and G.~H. Toumbev, J. Phys. G: Nucl. Phys. 8
  (1982) 997--1005.

\bibitem{Crespi2010299}
F.~C.~L. Crespi et~al., Nucl. Instr. Meth. A 620 (2010) 299 -- 304.

\bibitem{Recchia200960}
F.~Recchia et~al., Nucl. Instr. Meth. A 604 (2009) 60 -- 63.

\bibitem{Recchia2009555}
F.~Recchia et~al., Nucl. Instr. Meth. A 604 (2009) 555 -- 562.

\end{thebibliography}

\end{document}